\documentclass[manuscript,screen,nonacm]{acmart}
\usepackage{booktabs}  
\usepackage{graphicx}  
\usepackage{multirow}  
\usepackage{array}   
\usepackage{adjustbox}
\usepackage{makecell}
\usepackage{tikz}
\usetikzlibrary{positioning,arrows.meta,shapes.geometric,calc}

\usepackage{amssymb}

\AtBeginDocument{%
  }

\setcopyright{none}
\begin{document}

% \title{A Survey of Compound AI Systems}
\title{From Standalone LLMs to Integrated Intelligence: A Survey of Compound AI Systems}

% \author{Jiayi Chen}
% % \authornote{Both authors contributed equally to this research.}
% \email{jc2693@njit.edu}
% % \orcid{1234-5678-9012}
% % \author{G.K.M. Tobin}
% % \email{webmaster@marysville-ohio.com}
% \affiliation{%
%   \institution{New Jersey Institute of Technology}
%   \city{Newark}
%   \state{New Jersey}

% \author{Jiayi Chen}
% \email{jc2693@njit.edu}
% \affiliation{%
%   \institution{New Jersey Institute of Technology}
%   \city{Newark}
%   \state{New Jersey}

% \author{Junyi Ye}
% \email{jy394@njit.edu}
% \affiliation{%
%   \institution{New Jersey Institute of Technology}
%   \city{Newark}
%   \state{New Jersey}

% \author{Guiling Wang}
% \email{gwang@njit.edu}
% \affiliation{%
%   \institution{New Jersey Institute of Technology}
%   \city{Newark}
%   \state{New Jersey}

\author{Jiayi Chen}
\email{jc2693@njit.edu}

\author{Junyi Ye}
\email{jy394@njit.edu}

\author{Guiling Wang}
\email{gwang@njit.edu}

\affiliation{%
  \institution{New Jersey Institute of Technology}
  \city{Newark}
  \state{New Jersey}
  \country{USA}
}

\renewcommand{\shortauthors}{Chen et al.}

\begin{abstract}
Compound AI Systems (CAIS) are an emerging paradigm that integrates large language models (LLMs) with external components, including retrievers, agents, tools, and orchestrators, to overcome the limitations of standalone models in tasks requiring memory, reasoning, real-time grounding, and multimodal understanding. These systems enable more capable and context-aware behaviors by composing multiple specialized modules into cohesive workflows. Despite growing adoption in both academia and industry, the CAIS landscape remains fragmented and lacks a unified framework for analysis, taxonomy, and evaluation. In this survey, we define the concept of CAIS, propose a multi-dimensional taxonomy based on component roles and orchestration strategies, and analyze four foundational paradigms: Retrieval-Augmented Generation (RAG), LLM Agents, Multimodal LLMs (MLLMs), and Orchestration. We review representative systems, compare design trade-offs, and summarize evaluation methodologies across these paradigms. Finally, we identify key challenges---including scalability, interoperability, benchmarking, and coordination---and outline promising directions for future research. This survey aims to provide researchers and practitioners with a comprehensive foundation for understanding, developing, and advancing the next generation of system-level artificial intelligence.

\end{abstract}

%    <concept>
%        <concept_id>10010147.10010178.10010179</concept_id>
%        <concept_desc>Computing methodologies~Natural language processing</concept_desc>
%        <concept_significance>500</concept_significance>
%        </concept>
%    <concept>
%        <concept_id>10010147.10010257</concept_id>
%        <concept_desc>Computing methodologies~Machine learning</concept_desc>
%        <concept_significance>100</concept_significance>
%        </concept>
%  </ccs2012>

% \ccsdesc[500]{Computing methodologies~Artificial intelligence}
% \ccsdesc[500]{General and reference~Surveys and overviews}



\maketitle

\section{Introduction}
\label{sec:introduction}

Large Language Models (LLMs) based on the Transformer architecture \cite{vaswani2017attention} have rapidly evolved from academic prototypes to foundational infrastructure for modern artificial intelligence, supporting billion-user chat interfaces, enterprise copilots, scientific discovery tools, and automated code generation systems. Flagship LLM families from leading providers—including OpenAI's GPT series~\cite{achiam2023gpt}, Google's Gemini~\cite{team2023gemini}, and Anthropic's Claude~\cite{claude_language_model}—routinely surpass human baselines on reasoning and NLP benchmarks, and the global market for generative AI is projected to exceed \$1.3~trillion by 2032~\cite{bloomberg2023generative}.

However, the same properties that make LLMs compelling—massive pretraining on static corpora and autoregressive token prediction—also introduce structural limitations: \textbf{hallucination} (fluent but factually inaccurate output), \textbf{staleness} (inability to access post-training knowledge), and \textbf{bounded reasoning} (finite context windows that constrain multi-hop reasoning and long-horizon task decomposition). These limitations impede safe and effective deployment in dynamic, real-world environments that require recency, factual reliability, and compositional reasoning.

To overcome these barriers, the community is converging on a new systems paradigm: \textbf{Compound AI Systems (CAIS)}. We define CAIS as modular and extensible architectures that integrate LLMs with specialized external components, including high-recall retrievers, tool-using agents, symbolic planners, long-term memory modules, multimodal encoders, and orchestration frameworks, to perform complex, dynamic, and high-precision tasks. By decoupling sub-task responsibilities and intelligently routing them to appropriate modules, CAIS extends the capabilities of LLMs far beyond what any monolithic model can achieve.

Early deployments demonstrate the transformative potential of this paradigm. Retrieval-augmented assistants, such as Perplexity.ai, provide real-time answers with chain-of-thought citations \cite{perplexityAI}. GitHub Copilot-X orchestrates code reasoning, repository search, and test generation, increasing developer throughput by over 55\% \cite{peng2023impact}. 
In radiology, multimodal pipelines coupled with rule-based triage agents have reduced report turnaround by 30\% while maintaining expert-level accuracy \cite{radlogics2021ai}. 
These cases mark a shift in design philosophy: from LLMs as autonomous \emph{soloists} to \emph{conductors} orchestrating heterogeneous AI ensembles.

Despite growing interest, a systematic understanding of CAIS remains elusive. Recent literature has addressed isolated components of this ecosystem, including surveys on retrieval-augmented generation (RAG)~\cite{fan2024survey}, LLM-based agents~\cite{li2025review}, multi-agent frameworks~\cite{guo2024large}, and LLM-driven system optimization~\cite{lin2024llm}—but these efforts remain fragmented. Some concentrate on narrow aspects such as prompt engineering~\cite{ma2024large}, benchmark analysis~\cite{ferrag2025llm}, or agent communication protocols~\cite{yan2025beyond}, without addressing the architectural interactions and trade-offs across the entire CAIS stack. None provides a holistic, system-level synthesis.

In contrast, our survey provides the first comprehensive, cross-axis systems-level synthesis of Compound AI Systems. We integrate four foundational axes—retrieval, agency, multimodal perception, and orchestration—into a cohesive framework, identify recurring design patterns, trade-offs, and failure modes, and propose an evaluation paradigm that addresses factuality, efficiency, safety, and human-centered utility.

Our analysis is structured along four axes that reflect the core dimensions of CAIS. Specifically, we make the following contributions:

\begin{enumerate}
  \item \textbf{Multi-dimensional taxonomy.} We organize the CAIS landscape across four orthogonal axes—RAG, LLM Agents, MLLMs, and Orchestration—establishing a shared vocabulary for comparative study, grounded in a systematic review of over 220 papers (Section~\ref{sec:methodology}).

  \item \textbf{Unified pipeline model.} We introduce a cross-axis architecture model (Figure~\ref{fig:unified_pipeline}) and formal notation $f(L, C, D)$ that shows how the four axes interact in a complete end-to-end CAIS pipeline, addressing the absence of a unified conceptual framework in prior surveys.

  \item \textbf{Comparative analysis.} We provide four structured comparison tables—RAG retriever paradigms (Table~\ref{tab:rag_comparison}), LLM agent reasoning frameworks (Table~\ref{tab:agent_comparison}), MLLM architectures (Table~\ref{tab:mllm_comparison}), and orchestration frameworks (Table~\ref{tab:orchestration_comparison})—that distill design trade-offs and system-level implications across each axis.

  \item \textbf{Standardization coverage.} We survey the emerging interoperability landscape—including the Model Context Protocol (MCP), Agent-to-Agent (A2A), Agent Communication Protocol (ACP), Agent Network Protocol (ANP), and Agent Spec—as a cross-cutting architectural concern, structured as a three-layer stack from tool schemas through agent communication protocols.

  \item \textbf{Research agenda.} We synthesize open challenges in evaluation, component integration, standardization governance, and multimodal alignment, and outline promising future research directions.
\end{enumerate}

The remainder of this survey is organized as follows. Section~\ref{sec:related_work} positions this work against closely related surveys. Section~\ref{sec:methodology} describes our systematic literature search and inclusion criteria. Section~\ref{sec:architecture} formalizes the CAIS representation $f(L, C, D)$ and presents the unified cross-axis pipeline model. Sections~\ref{sec:rag}--\ref{sec:orchestration} examine each axis in depth with comparative analysis; Section~\ref{sec:orchestration} additionally covers the emerging standardization landscape. Section~\ref{sec:benchmarks_and_evaluation_metrics} reviews evaluation methodologies, Section~\ref{sec:challenges_limitations_and_opportunities} synthesizes open research directions, and Section~\ref{sec:conclusion} concludes with a forward-looking agenda.

\section{Related Work}
\label{sec:related_work}

A growing body of surveys addresses aspects of the Compound AI Systems landscape. Table~\ref{tab:survey_comparison} positions this survey against the most closely related prior work. We organize the discussion into three groups: CAIS-focused surveys, agent and multi-agent surveys, and evaluation-focused contributions.

\subsection{CAIS-Focused Surveys}

\textbf{LLM-based Optimization of Compound AI Systems}~\cite{lin2024llm} surveys methods for optimizing compound AI pipelines in which an LLM itself acts as the optimizer, covering prompt optimization, in-context learning, and pipeline-level training. Its scope is deliberately restricted to \textit{training-time, LLM-as-optimizer} approaches and explicitly excludes architecture search, topology optimization, and most multi-agent coordination strategies. It does not address RAG retriever design, multimodal integration, orchestration patterns, or system-level evaluation.

\textbf{Compound AI Systems Optimization: A Survey of Methods, Challenges, and Future Directions}~\cite{lee2025compound} broadens the optimization scope to include both numerical and natural-language learning signals, organizing 26 representative works into a 22-cell taxonomy. However, it remains focused on the optimization slice of the CAIS design space and does not provide a general architectural synthesis, a cross-axis taxonomy, or coverage of multimodal systems, standardization, or deployment-oriented trade-offs.

\textbf{Evaluating Compound AI Systems through Behaviors, Not Benchmarks}~\cite{bhagat2025evaluating} proposes a behavior-driven testing framework for information-seeking compound systems that generates scenario-based test specifications and selects diverse cases using submodular optimization. It is limited to retrieval-augmented conversational QA and does not cover the broader taxonomy, architectural design space, multimodal systems, or orchestration strategies.

\subsection{Agent and Multi-Agent Surveys}

\textbf{A Review of Prominent Paradigms for LLM-Based Agents}~\cite{li2025review} surveys tool use, planning, and feedback learning as agent construction paradigms; it treats RAG instrumentally rather than as an independent axis and omits multimodal perception and orchestration frameworks.

\textbf{Large Language Model Based Multi-Agents}~\cite{guo2024large} covers role assignment, communication protocols, and emergent behaviors in multi-agent LLM systems; it intersects with the Orchestration axis but omits RAG, multimodal LLMs, and cross-component evaluation.

\textbf{Beyond Self-Talk}~\cite{yan2025beyond} surveys agent communication mechanisms including MCP, A2A, and ANP, complementing the standardization discussion here but not addressing RAG, multimodal systems, or heterogeneous component orchestration.

\textbf{The Rise and Potential of LLM-Based Agents}~\cite{xi2025rise} provides a broad agent survey spanning perception, memory, planning, and action, but does not address compound architectures that integrate RAG, multimodal encoders, and orchestration as first-class dimensions.

\subsection{Positioning of This Survey}

\begin{table*}[t]
\centering
\caption{Comparison of this survey with closely related prior work. Checkmarks indicate substantive coverage; partial coverage is noted where applicable.}
\label{tab:survey_comparison}
\begin{adjustbox}{max width=\textwidth}
\begin{tabular}{lcccccccc}
\toprule
\textbf{Survey} & \textbf{Year} & \textbf{RAG} & \textbf{LLM Agents} & \textbf{MLLM} & \textbf{Orchestration} & \textbf{Evaluation} & \textbf{Standardization} & \textbf{Scope} \\
\midrule
Lin et al.~\cite{lin2024llm}               & 2024 & Partial & Partial & \texttimes & \texttimes & Partial & \texttimes & Optimization only \\
Lee et al.~\cite{lee2025compound}          & 2025 & Partial & \checkmark & \texttimes & Partial & \texttimes & \texttimes & Optimization only \\
Bhagat et al.~\cite{bhagat2025evaluating}  & 2025 & Partial & \texttimes & \texttimes & \texttimes & \checkmark & \texttimes & Info-seeking QA \\
Li~\cite{li2025review}                     & 2025 & Partial & \checkmark & \texttimes & \texttimes & Partial & \texttimes & Agent paradigms \\
Guo et al.~\cite{guo2024large}             & 2024 & \texttimes & \checkmark & \texttimes & Partial & Partial & \texttimes & Multi-agent \\
Yan et al.~\cite{yan2025beyond}            & 2025 & \texttimes & \checkmark & \texttimes & Partial & \texttimes & Partial & Communication \\
Xi et al.~\cite{xi2025rise}                & 2025 & Partial & \checkmark & Partial & \texttimes & Partial & \texttimes & Agent systems \\
\midrule
\textbf{This survey}                       & \textbf{2026} & \checkmark & \checkmark & \checkmark & \checkmark & \checkmark & \checkmark & \textbf{Full CAIS} \\
\bottomrule
\end{tabular}
\end{adjustbox}
\end{table*}

The core differentiator of this survey is its \textbf{systems-level synthesis} across all four foundational axes of CAIS simultaneously. Prior surveys either focus on a single axis (RAG, agents, or evaluation) or restrict their scope to optimization methods within compound systems. None provide a unified taxonomy that integrates retrieval, agency, multimodal perception, and orchestration into a single analytical framework, nor do they address the emerging standardization landscape (MCP, A2A, ACP, ANP) as a cross-cutting architectural concern. This survey fills that gap by offering the first comprehensive, cross-axis treatment of Compound AI Systems from architecture through evaluation to open research challenges.

\section{Survey Methodology}
\label{sec:methodology}

This section describes the systematic process used to conduct this survey, including the literature search strategy, inclusion and exclusion criteria, and the rationale for the scope boundaries that define the four axes of our taxonomy.

\subsection{Literature Search Strategy}

We conducted a structured literature search across four major scholarly databases: the ACM Digital Library, arXiv, IEEE Xplore, and Semantic Scholar. The search was performed between December 2024 and May 2026 and targeted publications from 2020 through early 2026, a window chosen to capture the period during which LLM-augmented systems emerged as a distinct research paradigm following the introduction of GPT-3~\cite{brown2020language} and the seminal RAG work of Lewis et al.~\cite{lewis2020retrieval}.

The primary search queries used were combinations of the following terms: \textit{``compound AI system''}, \textit{``retrieval-augmented generation''}, \textit{``RAG''}, \textit{``LLM agent''}, \textit{``tool-augmented language model''}, \textit{``multimodal large language model''}, \textit{``AI orchestration''}, \textit{``multi-agent LLM''}, and \textit{``LLM pipeline''}. Secondary searches were conducted using the reference lists of highly cited papers identified in the primary pass (snowballing). We supplemented automated searches with manual inspection of proceedings from ACL, EMNLP, NeurIPS, ICML, ICLR, CVPR, and ACM SIGKDD, as these venues have published a disproportionate share of CAIS-relevant work.

In total, we identified approximately 400 candidate papers. After applying the inclusion and exclusion criteria described below, over 220 papers were retained and are cited or discussed in this survey.

\subsection{Inclusion and Exclusion Criteria}

A paper was \textbf{included} if it: (1) used an LLM (at least one billion parameters, or a smaller model explicitly designed to replicate such capabilities) as a primary reasoning or generation component; (2) combined that LLM with at least one external module such as a retriever, tool, memory store, API, multimodal encoder, or orchestration framework; (3) presented novel empirical results, a new architecture, a benchmark, or a systematic comparative analysis; and (4) was publicly accessible via one of the searched databases or arXiv.

A paper was \textbf{excluded} if it focused exclusively on LLM pretraining or fine-tuning without external component integration (i.e., it refined the model itself rather than building a compound system), addressed non-Transformer or small language models without relevance to LLM-based compound architectures, was a purely theoretical contribution without empirical validation, was a duplicate or extended abstract of an already-included paper, or concerned only hardware infrastructure or low-level systems software without addressing AI component integration.

\subsection{Scope Boundaries and Axis Selection}

The four axes of our taxonomy---RAG, LLM Agents, Multimodal LLMs (MLLMs), and Orchestration---were selected based on two criteria: (1) they represent the dominant and most independently studied integration paradigms in the surveyed literature, and (2) together they cover the full spectrum of how LLMs are augmented with external capabilities in practice.

\textit{Retrieval-Augmented Generation (RAG)} was selected because grounding LLMs in external non-parametric memory is the most prevalent form of LLM augmentation, accounting for a large fraction of the surveyed literature and underpinning many downstream agent and multimodal systems.

\textit{LLM Agents} were selected because autonomous decision-making and tool use---where the LLM plans, acts, and reflects in iterative loops---constitute a qualitatively distinct capability class that extends beyond passive retrieval.

\textit{Multimodal LLMs (MLLMs)} were selected because extending LLMs to perceive and reason over non-textual modalities (images, audio, video) introduces a distinct integration challenge involving modality-specific encoders and cross-modal alignment that is not captured by the other axes.

\textit{Orchestration} was selected because the coordination of multiple LLMs and components into coherent pipelines---including topology, scheduling, memory management, and system objectives---is an independent design dimension that applies regardless of which specific components are integrated.

The following topics are explicitly \textit{out of scope}: LLM pretraining and continuous pretraining, parameter-efficient fine-tuning (LoRA, adapters) without external component integration, pure natural language processing benchmarks, hardware-level serving infrastructure, and AI safety alignment research that does not involve compound system architectures.

How this survey relates to and is differentiated from prior work on individual CAIS axes is discussed above in the Related Work section (Section~\ref{sec:related_work}).

\section{Architecture of Compound AI Systems}
\label{sec:architecture}

\subsection{Definition}
The term \emph{Compound AI Systems} (CAIS) first appeared in a post from Berkeley Artificial Intelligence Research (BAIR) \cite{compound-ai-blog}. In this post, the authors illustrate the trend where better AI performance is increasingly achieved through compound systems rather than standalone LLMs. In this survey, we provide a comprehensive definition: A CAIS is a framework that integrates LLMs, external components, and system-level designs. Its purpose is to address complex tasks that exceed the capabilities of standalone LLMs. It can be described using four dimensions: Retrieval-Augmented Generation (RAG), Multimodal LLMs (MLLMs), LLM Agents, and Orchestration. Pretraining, continuous pretraining, and fine-tuning without combining other components are not considered CAIS, as they refine the LLM itself rather than composing it with external components. 

\subsection{General Representation of Compound AI Systems}
From a general perspective, CAIS can be understood as a system comprising integrated interacting components and core LLMs. Components can be external tools, models, agents, multimodal encoders, prompt engineering techniques, and self-optimization. LLMs may include general-purpose models or fine-tuned models trained for specific tasks. Both components and LLMs are essential parts of CAIS.
As such, the general formulation of a compound AI system can be described as:

\begin{equation}
\text{Compound AI System} = f(L, C, D)
\end{equation}

\noindent
Where:
\begin{itemize}
    \item \( L = \{\text{LLM}_1, \text{LLM}_2, \dots, \text{LLM}_M\} \): The set of all LLMs in the system.
    \item \( C = \{C_1, C_2, \dots, C_N\} \): The set of all components, each providing functionality \( f_i \), output \( o_i \), and parameters \( p_i \).
    \item \( D \): The system design that defines the architecture and interactions between \( L \) and \( C \), including orchestration, topology, and other high-level design principles.
\end{itemize}

\noindent
The function \( f(L, C, D) \) abstracts the interactions of the core language models (\(L\)), the auxiliary modules that enhance or extend their functionality (\(C\)), and the overarching framework governing how they are connected and coordinated to achieve the desired outcomes (\(D\)).

\subsection{Dimensions of Compound AI Systems}
CAIS are organized along four axes: \textit{RAG} (grounding LLM outputs in externally retrieved context), \textit{MLLMs} (extending LLMs to perceive and reason over images, audio, and video via modality-specific encoders), \textit{LLM Agents} (autonomous reasoning, planning, and tool use in iterative loops), and \textit{Orchestration} (the structural and coordination design that integrates all components). Each axis is examined in depth in Sections~\ref{sec:rag}–\ref{sec:orchestration}.

\subsection{Cross-Axis Interactions in a Complete CAIS Pipeline}
\label{sec:cross_axis}

While the four axes—RAG, LLM Agents, MLLMs, and Orchestration—are analyzed independently in Sections~\ref{sec:rag}–\ref{sec:orchestration}, real-world Compound AI Systems routinely combine all four within a single pipeline. This subsection illustrates how the axes interact by tracing the flow of a representative end-to-end system: a \textit{multimodal research assistant} that accepts image-and-text queries and produces grounded, tool-augmented responses.

\begin{figure}[t]
    \centering
    \begin{tikzpicture}[
        box/.style={rectangle, draw, rounded corners=4pt,
                    minimum width=2.8cm, minimum height=0.9cm,
                    align=center, font=\small},
        arrow/.style={->, thick, >=stealth},
        lbl/.style={font=\footnotesize\itshape, inner sep=2pt}
    ]
    %% Row 1 (y=0): User Query  MLLM Encoder  RAG Retriever
    %% Column spacing 4.4cm gives arrows enough room for labels
    \node[box, fill=gray!15]   (user)  at (0,   0) {User Query\\(image + text)};
    \node[box, fill=blue!12]   (mllm)  at (4.4, 0) {MLLM Encoder\\(\S\ref{sec:multimodal})};
    \node[box, fill=green!12]  (rag)   at (8.8, 0) {RAG Retriever\\(\S\ref{sec:rag})};

    %% Row 2 (y=-2.2): LLM Agent  Orchestration  Final Response
    \node[box, fill=orange!12] (agent) at (8.8,-2.2) {LLM Agent\\(\S\ref{sec:llm_agent})};
    \node[box, fill=purple!12] (orch)  at (4.4,-2.2) {Orchestration\\(\S\ref{sec:orchestration})};
    \node[box, fill=gray!15]   (out)   at (0,  -2.2) {Final Response};

    %% Row-1 forward arrows
    \draw[arrow] (user)  -- node[above, lbl]{encode}     (mllm);
    \draw[arrow] (mllm)  -- node[above, lbl]{query vec.} (rag);

    %% Down connector: RAG -> LLM Agent, shifted right so it clears the dashed arrow
    \draw[arrow] ([xshift= 0.18cm]rag.south)
              -- node[right, lbl]{context}
                 ([xshift= 0.18cm]agent.north);

    %% Row-2 reverse arrows (right to left)
    \draw[arrow] (agent) -- node[above, lbl]{subtasks} (orch);
    \draw[arrow] (orch)  -- node[above, lbl]{result}   (out);

    %% Dashed feedback: Orchestration -> up into gap -> across -> up into RAG from below
    %% rag is 4.4cm right of orch; shift left 0.18cm -> horizontal run = 4.22cm
    \draw[arrow, dashed]
        (orch.north)
        -- ++(0, 0.65)
        -- node[above, lbl]{tool call / re-retrieve} ++(4.22, 0)
        -- ([xshift=-0.18cm]rag.south);
    \end{tikzpicture}
    \caption{Unified CAIS pipeline showing how the four axes interact end-to-end. Row~1 (left to right): a multimodal query is encoded by the MLLM component (\S\ref{sec:multimodal}) and the resulting vector drives the RAG Retriever (\S\ref{sec:rag}). Row~2 (right to left): retrieved context is passed to the LLM Agent (\S\ref{sec:llm_agent}), which delegates subtasks to the Orchestration layer (\S\ref{sec:orchestration}), which in turn synthesizes the final response. The dashed arc represents iterative feedback: the Orchestration layer may trigger additional retrieval or tool calls before the response is finalized.}
    \Description{Flow diagram showing a multimodal user query passing sequentially through MLLM encoding, RAG retrieval, LLM agent planning, and orchestration before producing a final response. A dashed arrow from the orchestration layer back to the RAG retriever indicates iterative feedback loops.}
    \label{fig:unified_pipeline}
\end{figure}

As illustrated in Figure~\ref{fig:unified_pipeline}, the MLLM component encodes the image query into token embeddings, the RAG retriever grounds the LLM in up-to-date external knowledge, the LLM Agent decomposes the task and invokes tools via a ReAct-style loop~\cite{yao2023react}, and the Orchestration layer schedules all component interactions—triggering additional retrieval when needed before synthesizing the final response. The four axes co-constrain one another: MLLM design determines encoder integration, RAG design governs retriever choice, the Agent axis defines the reasoning protocol, and Orchestration specifies the coordination topology $D$. The remainder of this survey examines each axis in depth, beginning with RAG in Section~\ref{sec:rag}.

\section{Retrieval-Augmented Generation (RAG)}
\label{sec:rag}

\subsection{Motivation and General Framework}

LLMs are highly capable but exhibit several limitations, including hallucinations \cite{huang2023survey}, outdated knowledge, unstable contextual understanding, and high computational costs associated with retraining or fine-tuning. RAG addresses these limitations by incorporating external documents, datasets, or search engines as non-parametric memory~\cite{lewis2020retrieval}.
This approach allows RAG to enhance the performance of LLMs in a cost-efficient manner without requiring extensive retraining on massive datasets. Additionally, RAG provides mechanisms for controlling the quality of the generated responses, thereby improving their reliability and relevance. Some popular RAG libraries, such as LangChain \cite{langchain}, Haystack \cite{haystack}, and LlamaIndex \cite{llama_index}, enable the efficient integration of retrieval with LLMs for scalable and reliable applications.

The RAG framework typically consists of two primary phases: Retrieval and Generation. In the retrieval phase, knowledge documents are segmented into smaller chunks, indexed, and represented as vectors through embedding techniques. Queries are similarly embedded, and the most relevant top-\textit{k} chunks are selected based on cosine similarity. In the generation phase, these retrieved documents serve as contextual input, which is combined with the original query and processed by the LLM. The LLM then generates a response that is both accurate and contextually grounded~\cite{lewis2020retrieval}.

\begin{figure}[t]
    \centering
    \includegraphics[width=0.82\linewidth]{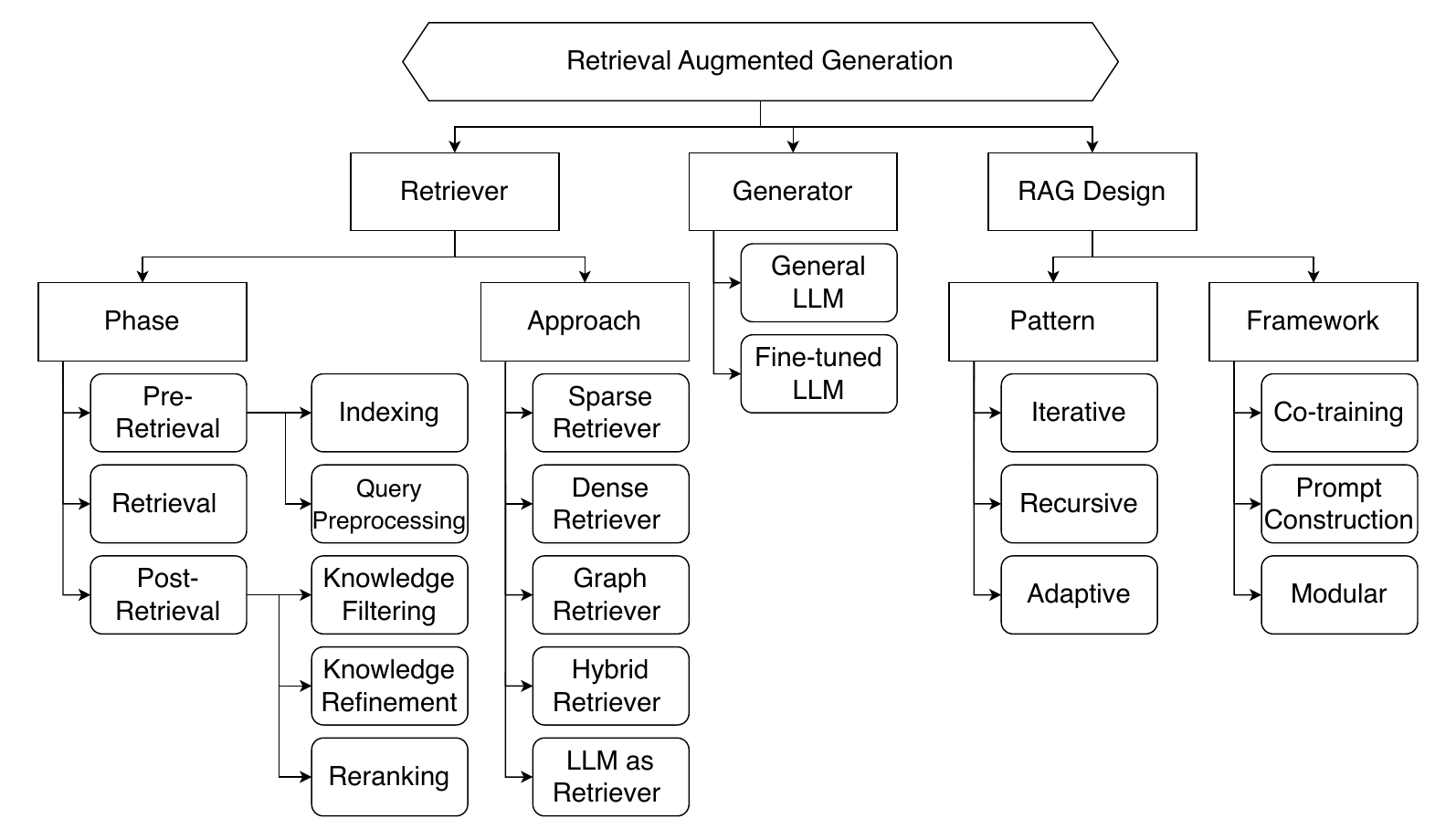}
    \caption{Taxonomy of Retrieval-Augmented Generation (RAG) systems. The diagram categorizes the core components into three primary modules: Retriever, Generator, and RAG design.}
    \Description{Diagram showing the three core modules of Retrieval-Augmented Generation (RAG) systems: Retriever, Generator, and RAG design, organized in a taxonomy.}
    \label{fig:RAG}
\end{figure}

This survey categorizes the RAG architecture into three key components: the \textit{Retriever}, the \textit{Generator}, and \textit{RAG Design}. The detailed architecture is illustrated in Figure~\ref{fig:RAG}. As shown, the Retriever identifies and retrieves relevant documents from external databases or search engines through various methods. The Generator processes the retrieved content to produce retrieval-augmented answers, utilizing either pretrained or fine-tuned large language models (LLMs). The RAG design encompasses the overall orchestration of the system, characterized by distinct patterns or frameworks.

\subsection{Retriever}

The primary role of the retriever in RAG systems is to retrieve information and external knowledge. 
It is essential that the retriever effectively identifies and retrieves relevant context to support accurate and meaningful responses. The retriever in RAG operates along two dimensions: \textit{Phase} and \textit{Approach}.
With regard to \textit{Phase}, the retriever can be categorized into three phases: \textit{Pre-Retrieval}, \textit{Retrieval}, and \textit{Post-Retrieval}, based on the sequence of the retrieval process. 

\textit{Pre-Retrieval} refers to the process of converting external knowledge into vector representations, including query preprocessing, chunking, embedding, and indexing \cite{multimodal_chunking_rag, datastax_rag_indexing}. Query preprocessing improves retrieval quality through techniques such as query rewriting, where a trainable rewriter modifies input queries for better retrieval~\cite{ma2023query}, and docid-based retrieval, where models like DSI map document text directly to unique identifiers for efficient lookup~\cite{tay2022transformer}.

\textit{Retrieval} involves locating documents relevant to a query by measuring similarity. For example, Yu et al.~\cite{yu2022automated} introduced an assertion retrieval method leveraging Jaccard similarity, while KAPING employs semantic similarity to retrieve relevant triples from knowledge graphs~\cite{baek2023knowledge}.

\textit{Post-Retrieval} focuses on refining initial retrieval results through filtering, refinement, and reranking. Knowledge filtering reduces the search space by applying specific constraints~\cite{dai2022promptagator, wang2023learning}, while knowledge refinement improves coherence and usability of retrieved information~\cite{xu2023recomp}. Reranking reorders and reassesses retrieved results to maximize relevance~\cite{lazaridou2022internet, glass2022re2g, ram2023context}.

From a functional perspective, retrievers can be classified into five types: \textit{Sparse Retriever}, \textit{Dense Retriever}, \textit{Graph Retriever}, \textit{Hybrid Retriever}, and \textit{LLM as Retriever}.
\textit{Sparse Retriever} uses sparse representations of text (e.g., BM25) based on explicit term matching between queries and documents~\cite{izacard2020leveraging}. \textit{Dense Retriever} leverages dense vector representations (e.g., BERT) to capture semantic similarity between queries and documents~\cite{lin2023train, yang2024leandojo, liu2023webglm}. \textit{Graph Retriever} utilizes graph structures (e.g., knowledge graphs) to locate relevant information by traversing nodes and edges~\cite{edge2024local, gaur2022iseeq, baek2023knowledge}. \textit{Hybrid Retriever} combines both sparse and dense retrieval approaches to leverage the strengths of explicit term matching and semantic similarity~\cite{lu2022reacc, glass2022re2g, baek2023knowledge}. \textit{LLM as Retriever} involves the use of LLMs to directly retrieve relevant knowledge based on input queries~\cite{ma2024fine}.

\subsection{Generator}
The Generator in RAG systems is essentially an LLM. It can be an original pretrained language model, such as T5~\cite{raffel2020exploring}, FLAN~\cite{wei2021finetuned} and LLaMA~\cite{touvron2023llama} (since succeeded by LLaMA~4~\cite{metaLlama4}), or a black-box pretrained language model, such as GPT-3~\cite{brown2020language}, GPT-4~\cite{achiam2023gpt}, Gemini~\cite{team2023gemini}, and Claude~\cite{claude_language_model}. Alternatively, the generator can also be a fine-tuned language model specifically tailored for a particular task. For instance, BART~\cite{lewis2020retrieval} and T5~\cite{izacard2023atlas} are fine-tuned alongside the retriever, a process commonly referred to as co-training or dual fine-tuning, to enhance the quality and consistency of retrieval~\cite{lin2023ra}. In other scenarios, the generator is fine-tuned to effectively filter retrieved results, retaining only relevant documents and discarding irrelevant ones~\cite{luo2023sail, yoran2023making, zhang2024raft}. Furthermore, the Generator can be trained and fine-tuned within a reinforcement learning framework to optimize performance in specific contexts~\cite{gaur2022iseeq}.

\subsection{RAG Design}
RAG design is characterized by two key dimensions: \textit{pattern} and \textit{framework}. The \textit{pattern} dimension describes how RAG systems retrieve and generate answers, encompassing iterative, recursive, and adaptive approaches. The \textit{framework} dimension refers to the structural framework of RAG design, including co-training, prompt construction, and modularity.

\textit{Iterative} is a design pattern that improves RAG systems through repeated cycles of retrieval and generation to refine outputs incrementally. For example, ITER-RETGEN~\cite{shao2023enhancing} utilizes outputs from previous iterations to refine both retrieval and generation, effectively addressing multi-hop reasoning and complex queries~\cite{khattab2022demonstrate, peng2023check, zhang2023repocoder, cheng2024lift}.

\textit{Recursive} is a design pattern in which retrieval and generation are applied in a nested manner to handle complex queries by breaking them down into simpler ones. For example, IRCoT~\cite{trivedi2022interleaving} combines retrieval with iterative chain-of-thought reasoning for knowledge-intensive, multi-step question answering~\cite{zhang2024raft, wang2023augmenting}.

\textit{Adaptive} is a design pattern that dynamically adjusts retrieval or generation strategies based on context or feedback. For example, SELF-RAG~\cite{asai2023self} enables LLMs to decide dynamically when to retrieve information, generate responses, and critique their outputs via self-reflection~\cite{jiang2023active, yu2023chain}.

\textit{Co-training} is a framework where the retriever and generator are jointly trained to enhance their collaboration. For example, RAG-end2end~\cite{siriwardhana2023improving} updates all RAG components asynchronously during training, including the retriever, generator, and external knowledge base encodings.

\textit{Prompt Construction} is a framework that focuses on designing and optimizing prompts to enable the generator to efficiently utilize retrieved information. For example, Ren et al.~\cite{ren2023investigating} present prompting strategies such as a priori judgment (evaluating a question before answering) and a posteriori judgment (assessing the correctness of an answer) to explore the impact of retrieval augmentation.

\textit{Modular} is a framework in which RAG itself is designed as independent modules, allowing for flexibility, easy replacement, and integration of different components. For example, REPLUG~\cite{shi2023replug} augments LLMs by incorporating external retrieval systems without altering the LLM's internal parameters~\cite{goyal2022retrieval, yu2023augmentation, chen2022decoupling}.

\subsection{Comparative Analysis of RAG Retriever Approaches}
\label{sec:rag_comparison}

The four principal retrieval paradigms—sparse, dense, graph-based, and hybrid—differ substantially in their trade-off profiles. Table~\ref{tab:rag_comparison} summarizes these differences across five dimensions that matter most for system design: retrieval latency, recall quality, scalability to large corpora, interpretability of retrieved evidence, and suitability for specialized domains.

\begin{table*}[t]
\centering
\renewcommand{\arraystretch}{1.2}
\caption{Comparative analysis of RAG retriever paradigms across key system-design dimensions.}
\label{tab:rag_comparison}
\begin{adjustbox}{max width=\textwidth}
\begin{tabular}{lccccc}
\toprule
\textbf{Retriever Type} & \textbf{Latency} & \textbf{Recall Quality} & \textbf{Scalability} & \textbf{Interpretability} & \textbf{Domain Specificity} \\
\midrule
Sparse (BM25, TF-IDF)~\cite{robertson2009probabilistic}   & Very low   & Moderate (lexical match) & Very high (inverted index) & High (term weights visible) & Low (no semantic alignment) \\
Dense (DPR, ANCE)~\cite{karpukhin2020dense,xiong2020approximate}     & Low--medium & High (semantic match)    & High (ANN index)           & Low (opaque embeddings)     & High (fine-tunable)        \\
Graph-based (GraphRAG)~\cite{edge2024local} & Medium--high & Very high (relational)  & Medium (graph overhead)    & Medium (edge-based)         & Very high (structured KBs) \\
Hybrid (BM25 + Dense)~\cite{luan2021sparse,glass2022re2g}  & Medium     & Highest (complementary)  & High                       & Medium                      & High                       \\
LLM-as-Retriever~\cite{tang2023llmknowledgegraph} & High & Medium--high (generative) & Low (inference cost)       & Low                         & Adapts to prompt            \\
\bottomrule
\end{tabular}
\end{adjustbox}
\end{table*}

\textbf{Key trade-offs.} Sparse retrievers are the most computationally efficient and fully interpretable but fail on vocabulary mismatch and cannot capture semantic similarity. Dense retrievers address this through learned embeddings at the cost of interpretability and the need for in-domain fine-tuning data. Graph-based retrieval excels when the knowledge base has explicit relational structure (e.g., knowledge graphs or code repositories) but incurs higher construction and query overhead. Hybrid retrievers combine the complementary strengths of sparse and dense approaches and consistently achieve the highest recall on benchmarks such as BEIR~\cite{thakur2021beir}, making them the preferred choice for general-purpose CAIS deployments where recall is critical. LLM-as-retriever approaches are the most flexible but carry the highest inference cost and are best suited to low-volume, high-precision scenarios where retrieval and reasoning are tightly coupled.

\subsection{Limitations and Future Trends}
\label{sec:rag_limitations}
Although RAG systems have developed rapidly with numerous innovative designs, modern RAG systems still face several limitations. For instance, seamlessly integrating retrieved information into the generation process remains a significant challenge, particularly for longer contexts or multimodal tasks~\cite{cheng2024lift}. Additionally, issues such as retrieval quality~\cite{cuconasu2024power}, scalability~\cite{liu2023tcra}, and knowledge conflicts or inconsistencies~\cite{trivedi2022interleaving} continue to hinder the effectiveness of modern RAG systems.

Promising directions include ready-to-use domain-specific RAG frameworks~\cite{xu2023tool, savelka2023explaining, yang2024leandojo, tsai2024rtlfixer}, deeper multimodal integration across text, images, and video~\cite{chen2022re, yasunaga2022retrieval, 9879687}, RAG combined with reinforcement learning for improved control~\cite{goyal2022retrieval, ma2023query}, and end-to-end LLM-as-retriever pipelines that supersede traditional retrieval-then-rank approaches~\cite{tay2022transformer, ma2024fine}.

Having surveyed the retrieval axis of CAIS, the next section turns to the agency axis: LLM Agents that reason, plan, and take actions through iterative interaction with tools and environments.

\section{LLM Agents}
\label{sec:llm_agent}

\begin{figure}[h]
    \centering
    \includegraphics[width=0.82\linewidth]{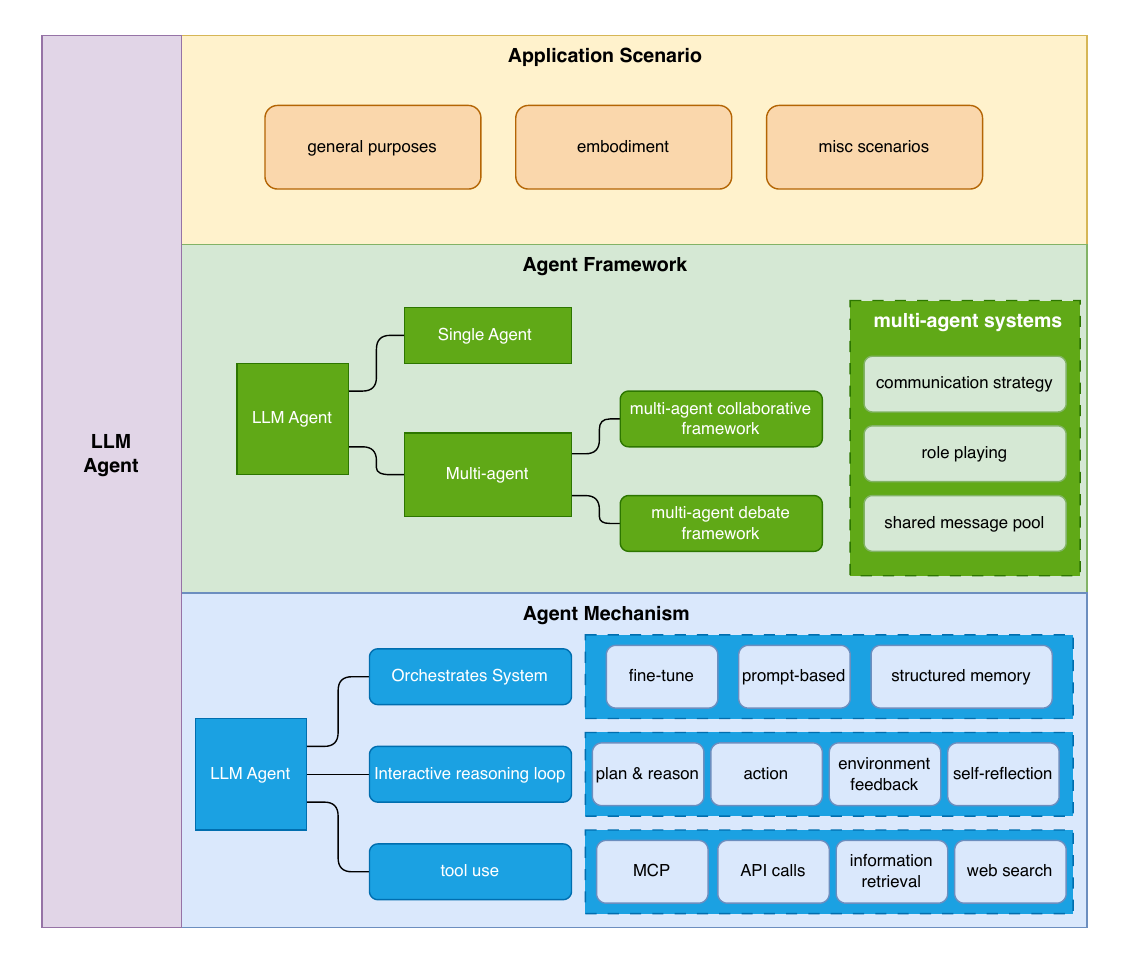}
    \caption{A structured overview of LLM agents across three dimensions: application scenarios (e.g., general-purpose, embodied), agent frameworks (single-agent and multi-agent architectures), and agent mechanisms (system orchestration, reasoning loops, and tool use).}
    \Description{Diagram presenting LLM agents along three axes: application scenarios like general-purpose and embodied agents; agent frameworks including single- and multi-agent architectures; and agent mechanisms such as orchestration, reasoning, and tool usage.}
    \label{fig:llm_agent}
    % \vspace{-0.4cm}
\end{figure}

LLM agents have emerged as a powerful paradigm for enabling intelligent, autonomous behavior.
The ecosystem of LLM agents can be broadly categorized into three interconnected layers: Application Scenario, Agent Framework, and Agent Mechanism, as illustrated in Figure \ref{fig:llm_agent}. 
Together, these layers define the operational architecture and capabilities of LLM agents, guiding both their theoretical design and practical implementation.

\subsection{Application Scenario}
LLM agents are applied across a spectrum of real-world and simulated environments. \textit{General-purpose agents} handle diverse tasks without domain-specific customization: Gato~\cite{reed2022generalist} uses a single set of weights to perform over 600 tasks across multiple modalities and embodiments, while MINEDOJO~\cite{fan2022minedojo} trains generalist embodied agents on an internet-scale Minecraft knowledge base. \textit{Embodied agents} are situated in physical or virtual environments: Inner Monologue~\cite{huang2022inner} enables closed-loop language feedback for robotic planning without additional training, and Voyager~\cite{wang2023voyager} achieves open-ended lifelong learning in Minecraft using an automatic curriculum and a code-based skill library. \textit{Miscellaneous scenarios} extend LLM agents to specialized domains, including reward function design for RL tasks (EUREKA~\cite{ma2023eureka}), knowledge-driven autonomous driving~\cite{wen2023dilu}, and end-to-end scientific experiment planning~\cite{boiko2023emergent}.

\subsection{Agent Framework}
The agent framework defines how LLM agents are organized, instantiated, and deployed. This layer encompasses the majority of paradigms in multi-agent frameworks, including multi-agent collaborative frameworks, multi-agent debate frameworks, and the workflow of multi-agent systems.

\subsubsection{Multi-Agent Collaborative Framework}
A multi-agent collaborative framework enables multiple LLM agents to work together toward a shared objective, often leveraging role specialization, communication protocols, and coordinated planning.
For example, MetaGPT \cite{hong2023metagpt} is a multi-agent collaborative framework for software development that integrates Standardized Operating Procedures (SOPs) into LLM-based agent workflows to improve coherence and accuracy. 
Moreover, AgentVerse \cite{chen2023agentverse} is a dynamic multi-agent collaboration framework inspired by human group problem-solving, where agents can adjust their roles, communicate, and collaborate across various tasks, including software development, consulting, and gaming. AgentVerse models the problem-solving process as a loop of four stages: expert recruitment, collaborative decision-making, action execution, and evaluation, allowing dynamic team adjustment based on feedback.
Another example worth mentioning is Talebirad et al.~\cite{talebirad2023multi}, where the authors propose a graph-based multi-agent collaboration framework in which agents and plugins form nodes with defined communication channels, allowing role-specific LLM agents to collaborate, dynamically generate new agents, provide feedback, and manage execution.

\subsubsection{Multi-Agent Debate Framework}
In a multi-agent debate framework, agents are intentionally assigned to argumentative roles, utilizing structured dialogue to explore divergent viewpoints, validate reasoning, or reach consensus through deliberation.
For instance, Du et al. \cite{du2023improving} propose a multi-agent debate framework in which multiple instances of language models collaboratively reason, critique, and revise their answers to enhance factuality and reasoning across various tasks. The authors design a system where several LLM agents independently generate responses, then iteratively review and revise their answers based on peer responses through multiple rounds of debate, using only prompting and without model fine-tuning. 
Furthermore, the Multi-Agent Debate (MAD) framework \cite{liang2023encouraging} involves two LLM-based agents (affirmative and negative) debating a topic iteratively with a judge LLM that decides when to stop and which response is correct, enabling exploration of diverse reasoning paths. 

\subsubsection{Multi-Agent Systems}
Multi-agent systems encompass any architecture involving two or more autonomous agents—LLMs that interact within a shared environment to achieve individual or collective goals.
\textit{Communication strategy} defines the protocols and methods through which LLM agents exchange information, coordinate actions, and negotiate meanings within a multi-agent setting. For example, ChatEval \cite{chan2023chateval} is a multi-agent debate framework that uses multiple LLMs with diverse roles to collaboratively evaluate generated text, simulating the quality and depth of human evaluation. ChatEval enables multiple LLM agents, each with a distinct role prompt (e.g., critic, scientist), to engage in structured discussion through designed communication strategies (one-by-one, simultaneous, or summarization-enhanced) before aggregating a final evaluation via majority vote or averaging. 

\textit{Role playing} assigns specific identities or perspectives to each agent, enabling structured interactions that reflect diverse viewpoints or specialized expertise. AutoGen~\cite{wu2023autogen} provides a general-purpose framework for multi-agent conversations using programmable chat patterns, while CAMEL~\cite{li2023camel} enables autonomous role-playing cooperation to generate high-quality conversational datasets.

A \textit{shared message pool} provides a centralized hub for asynchronous agent communication. MetaGPT~\cite{hong2023metagpt} implements this via a publish-subscribe model in which agents post and consume typed messages (e.g., Requirement, Design, Task, Code), enabling modular coordination without direct peer-to-peer dialogue.

\subsection{Agent Mechanism}

This layer covers the mechanisms of the LLM-driven agent: how the system is organized, how the LLM agent reasons, plans, perceives feedback from the environment, takes action, and reflects on failures. Moreover, tool use, one of the most essential features of the LLM agent, is also discussed in this layer. 
\subsubsection{Orchestration System}
An orchestration system coordinates the components and execution flow of an LLM agent, managing memory, tools, and reasoning steps to ensure coherent and goal-directed behavior. Most LLM agent systems are \textit{prompt-based}, which means the agent system guides the agent's behavior and reasoning by structuring task instructions, examples, and context directly within the input prompt to the language model, without modifying its internal parameters. Inner Monologue \cite{huang2022inner}, Voyager \cite{wang2023voyager}, and ChatEval \cite{chan2023chateval} are some representative examples that leverage the prompt-based method to compose their agent systems. However, there are agent systems that apply \textit{fine-tuning} as well. For instance, GPT4Tools \cite{yang2023gpt4tools} is a method that enables open-source language models to use multimodal tools by generating a tool-usage instruction dataset via self-instruction from GPT-3.5 and fine-tuning with LoRA. The authors fine-tune models, such as Vicuna-13B, using LoRA for efficient parameter adaptation. 

\textit{Structured memory management} is an essential component of the LLM agent system. Based on different use cases, memory storage typically employs a combination of short-term memory, long-term memory, and, in some instances, episodic memory. For example, CoALA \cite{sumers2023cognitive} is a conceptual framework inspired by cognitive science to organize and design modular, memory-augmented LLM-based agents, where it introduces a modular agent architecture with working and long-term memory (episodic, semantic, procedural). 
Moreover, ExpeL \cite{zhao2024expel} utilizes dual memory systems, where the agent autonomously collects successful and failed task trajectories, extracts reusable natural language insights, and retrieves relevant past examples to augment its decision-making during evaluation, all without modifying LLM weights. 
MemFlow~\cite{chen2026memflow} further frames memory access as an intent-driven orchestration problem for small language model agents, routing each query to specialized memory operations with tier-aware evidence packing and grounding-based escalation.

\subsubsection{Interactive Reasoning Loop}
The interactive reasoning loop is the iterative process through which an LLM agent perceives input, plans actions, executes tasks, and incorporates feedback to refine decisions over time.

\textit{Planning and reasoning:} ReAct~\cite{yao2023react} interleaves verbal reasoning traces with environment-changing actions via few-shot prompting, enabling dynamic replanning based on observed feedback. Tree of Thoughts~\cite{yao2023tree} extends this to tree-structured search over intermediate reasoning steps, supporting backtracking and breadth/depth-first exploration.

\textit{Action} is the execution step at which the agent issues a command—calling a tool, generating output, or interacting with an environment—without requiring explicit human instruction for each step.

\textit{Environment feedback:} Voyager~\cite{wang2023voyager} refines actions through an iterative prompting loop using execution errors and self-verification; LLM-Planner~\cite{song2023llm} dynamically updates task plans based on grounded environmental state.

\textit{Self-reflection:} Reflexion~\cite{shinn2023reflexion} stores verbal self-evaluations in episodic memory, allowing agents to improve across trials without weight updates. RCI~\cite{kim2023language} applies iterative critique-and-improvement across task, state, and agent grounding stages.

\subsubsection{Tool Use} Tool use empowers LLM agents to go beyond language generation by interacting with external services—such as APIs, computational tools, or information sources—to perform complex tasks like data processing, real-time querying, and integration with third-party applications.

\textit{Model Context Protocol} (MCP) \cite{anthropic2024mcp} is a standardized interface proposed by Anthropic that enables language models to interact with external tools, memory, documents, and user interfaces in a structured and flexible manner. It defines how models receive context and respond with function calls or natural language, facilitating advanced agent behaviors. For example, MCP servers such as AWS KB Retrieval facilitate retrieval from the AWS Knowledge Base using Bedrock Agent Runtime. A server like Brave Search enables web and local search capabilities using Brave's Search API. Moreover, a server like Filesystem provides secure file operations with configurable access controls.

LLM agents can invoke external services via \textit{API Calls}, enabling access to dynamic functionalities such as data processing, computation, or third-party applications. For example, ToolLLM \cite{qin2023toolllm} is a framework that empowers LLMs to effectively use over 16,000 real-world APIs by combining a new dataset (ToolBench), a decision-making algorithm (DFSDT), and a trained model (ToolLLaMA) to achieve robust tool-use capabilities. 

\textit{Information retrieval} allows agents to access relevant documents or data from structured or unstructured sources to support grounded reasoning and informed responses.
For instance, Sun et al.\ \cite{sun2023chatgpt} explored whether LLMs like ChatGPT and GPT-4 can serve as effective passage re-ranking agents in information retrieval and introduced a novel permutation generation method that allows LLMs to directly output ranked passage lists. 

\textit{Web search} enables agents to query the internet in real-time, expanding their knowledge beyond pretraining and allowing them to answer up-to-date or domain-specific questions.
For example, MIND2WEB \cite{deng2023mind2web} proposes MINDACT, a two-stage framework combining a small LM for webpage element ranking and a large LM for action prediction in a multiple-choice QA format, enabling the model to operate efficiently on long and noisy web pages. 

\subsection{Comparative Analysis of LLM Agent Reasoning Frameworks}
\label{sec:agent_comparison}

A diverse set of reasoning frameworks has emerged for LLM agents, each making distinct design choices about how the agent plans, invokes tools, recovers from errors, and incurs computational cost. Table~\ref{tab:agent_comparison} provides a structured comparison of the five most widely adopted frameworks.

% \begin{table*}[t]
% \centering
% \caption{Comparative analysis of LLM agent reasoning frameworks across key design dimensions.}
% \label{tab:agent_comparison}
% \begin{adjustbox}{max width=\textwidth}
% \begin{tabular}{lccccc}
% \toprule
% \textbf{Framework} & \textbf{Planning Depth} & \textbf{Tool Use} & \textbf{Error Recovery} & \textbf{Compute Cost} & \textbf{Representative Benchmark} \\
% \midrule
% Chain-of-Thought (CoT)~\cite{wei2022chain} & Medium (linear) & None (reasoning only) & None & Low (single pass)      & GSM8K, MATH \\
% ReAct~\cite{yao2023react}                  & Medium (interleaved) & Yes (action--observe loop) & Limited (via replanning) & Medium                & HotpotQA, ALFWorld \\
% Tree of Thoughts (ToT)~\cite{yao2023tree}  & High (branching)  & Optional              & High (backtracking)  & High (tree search)    & Game of 24, Creative Writing \\
% Reflexion~\cite{shinn2023reflexion}        & Medium            & Yes                   & High (verbal feedback + memory) & Medium--high  & HumanEval, AlfWorld \\
% Self-RAG~\cite{asai2023self}               & Low--medium        & Yes (retrieval-gated) & Medium (critique tokens) & Medium              & ASQA, TriviaQA, PopQA \\
% \bottomrule
% \end{tabular}
% \end{adjustbox}
% \end{table*}

\begin{table*}[t]
\centering
\renewcommand{\arraystretch}{1.2}
\caption{Comparative analysis of LLM agent reasoning frameworks across key design dimensions.}
\label{tab:agent_comparison}
\begin{adjustbox}{max width=\textwidth}
\begin{tabular}{lccccc}
\toprule
\textbf{Framework} & \textbf{Planning Depth} & \textbf{Tool Use} & \textbf{Error Recovery} & \textbf{Compute Cost} & \textbf{Representative Benchmark} \\
\midrule
Chain-of-Thought (CoT)~\cite{wei2022chain} & Medium (linear) & None (reasoning only) & None & Low (single pass)      & GSM8K~\cite{cobbe2021gsm8k}, MATH~\cite{hendrycks2021math} \\
ReAct~\cite{yao2023react}                  & Medium (interleaved) & Yes (action--observe loop) & Limited (via replanning) & Medium                & HotpotQA~\cite{yang2018hotpotqa}, ALFWorld~\cite{shridhar2020alfworld} \\
Tree of Thoughts (ToT)~\cite{yao2023tree}  & High (branching)  & Optional              & High (backtracking)  & High (tree search)    & Game of 24, Creative Writing \\
Reflexion~\cite{shinn2023reflexion}        & Medium            & Yes                   & High (verbal feedback + memory) & Medium--high  & HumanEval~\cite{chen2021humaneval}, AlfWorld~\cite{shridhar2020alfworld} \\
Self-RAG~\cite{asai2023self}               & Low--medium        & Yes (retrieval-gated) & Medium (critique tokens) & Medium              & ASQA~\cite{stelmakh2022asqa}, TriviaQA~\cite{joshi2017triviaqa}, PopQA~\cite{mallen2022not} \\
\bottomrule
\end{tabular}
\end{adjustbox}
\end{table*}

\textbf{Key trade-offs.} Chain-of-Thought reasoning is the lowest-cost option and effective for well-structured tasks but has no recovery mechanism and no tool access. ReAct introduces an action--observation loop that enables real-time tool use at modest additional cost, making it the most widely adopted baseline for tool-using agents. Tree of Thoughts achieves the deepest planning through explicit tree search and backtracking but incurs a multiplicative compute overhead that limits practical deployment to short horizons. Reflexion combines tool use with episodic memory and verbal self-reflection to achieve strong error recovery without weight updates, at the cost of multi-trial inference. Self-RAG integrates retrieval directly into the reasoning chain via learned critique tokens, achieving competitive factual accuracy with smaller models, but its gating mechanism requires supervised fine-tuning. For CAIS deployments, ReAct and Self-RAG represent the most practical choices, with Reflexion preferred when multi-turn error correction is critical.

Building on these frameworks, the next section examines how CAIS extend beyond text to incorporate multimodal perception through Multimodal LLMs.

\section{Multimodal Large Language Models (MLLMs)}
\label{sec:multimodal}

MLLMs extend the capabilities of LLMs beyond text-based understanding and generation to multimodal comprehension and content creation. These models are designed to process and generate information across multiple modalities, including text, images, audio, and video. By incorporating cross-modal learning, MLLMs enhance traditional LLMs, enabling them to integrate and reason across different types of data. 

This section categorizes MLLMs into four aspects: architecture of MLLMs, fusion strategy inside MLLMs, MLLMs' modality approaches, and tasks of MLLMs. From an architecture standpoint, an MLLM has four major components: Encoder, Visual Projector, Fusion Module, and Core LLM Model. Fusion strategy is another key aspect of MLLMs; different MLLMs employ various fusion strategies, broadly classified into early fusion, late fusion, cross-modal fusion, and hybrid fusion. In terms of modality, MLLMs can be further classified by their focused modality approaches: image, audio, video, or text-rich image. Moreover, based on different tasks and usages, MLLMs can be defined as multimodal or integrated with other techniques, such as MLLMs with Chain-of-Thought (CoT) or MLLMs with RAG.

\begin{figure}[t]
    \centering
    \includegraphics[width=0.70\linewidth]{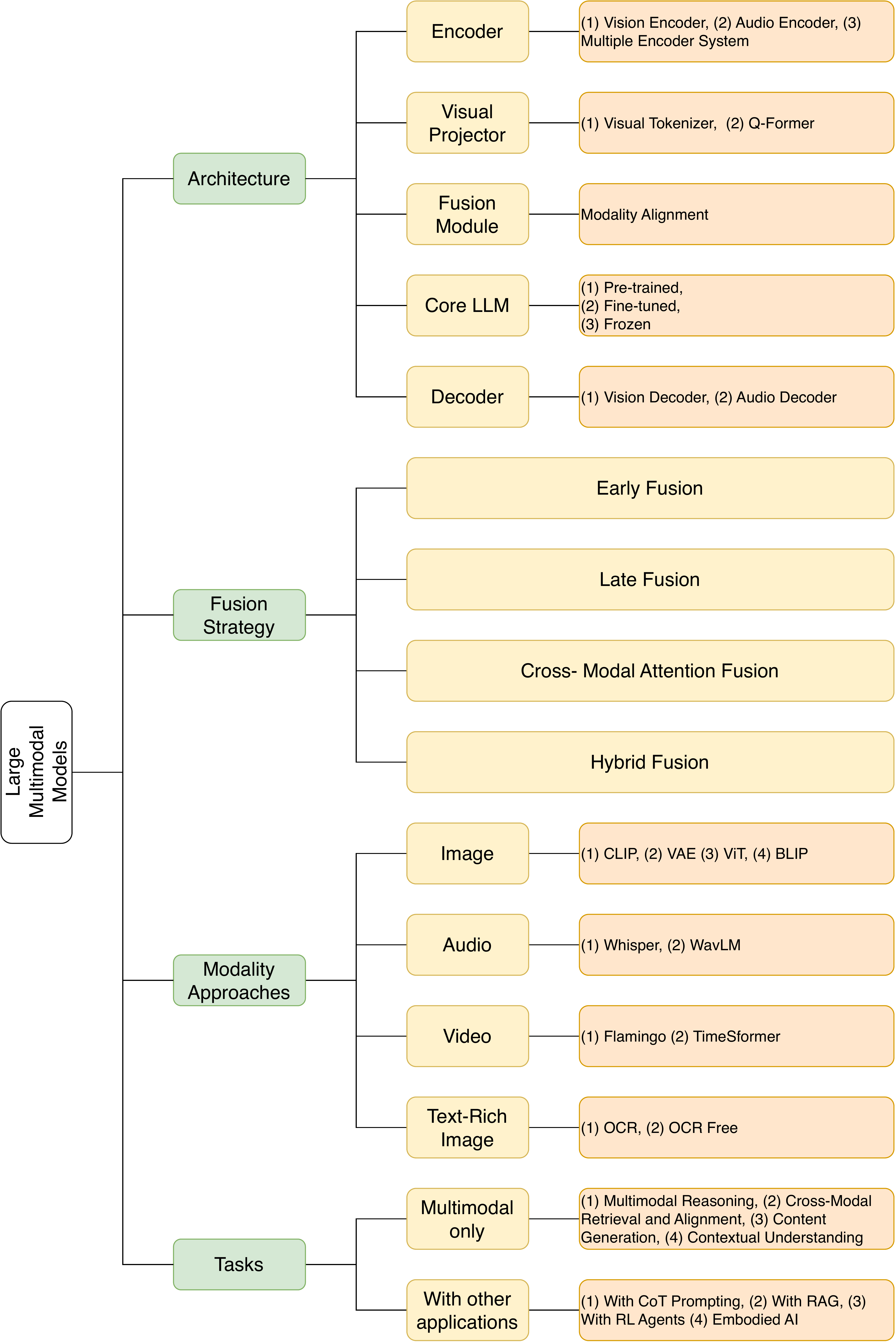}
    \caption{Overview of Multimodal Large Language Models. The diagram categorizes MLLMs by architecture, fusion strategy, modality, and tasks.}
    \Description{Diagram illustrating Multimodal Large Language Models (MLLMs), categorized by their architecture (e.g., encoder-decoder, unified models), fusion strategies (early, late, hybrid), supported modalities (such as text, image, video), and typical tasks.}
    \label{fig:multimodal}
\end{figure}

\subsection{Architecture}

An MLLM comprises four main components. The \textbf{Encoder} extracts modality-specific representations: a vision encoder (e.g., CLIP-based in LLaVA~\cite{liu2024visual}) converts pixel data into embeddings; an audio encoder (e.g., Whisper + BEATs in SALMONN~\cite{tang2023salmonn}) captures temporal and spectral features; and multi-encoder systems such as GLaMM~\cite{rasheed2024glamm} combine global and region-level visual encoders for grounded dialogue. The \textbf{Visual Projector} maps encoded features into the language model's token space—BLIP-2's Q-Former~\cite{li2023blip} uses learnable query vectors and cross-attention to compress visual representations into a fixed-length bottleneck, while ShareGPT4V~\cite{chen2025sharegpt4v} demonstrates that caption quality at this alignment stage critically determines downstream performance. The \textbf{Fusion Module} merges representations across modalities into a unified context. The \textbf{Core LLM} generates outputs from the fused representation; it may be used as-is (PaLM-E~\cite{driess2023palm} trains end-to-end with multimodal inputs), fine-tuned via LoRA adapters (MultiModal-GPT~\cite{gong2023multimodal}), or kept fully frozen while only connector weights are updated (FROMAGe~\cite{koh2023grounding}). Some models additionally include a \textbf{Decoder} for generative outputs across modalities (e.g., Emu2~\cite{sun2024generative} produces interleaved image-and-text via a visual decoder).

\begin{table}[t]
\centering
\renewcommand{\arraystretch}{1.2}
\caption{Representative encoder approaches for each modality in MLLMs.}
\label{tab:modality_approaches}
\begin{adjustbox}{max width=\linewidth}
\begin{tabular}{lllp{3.5cm}}
\toprule
\textbf{Model} & \textbf{Modality} & \textbf{Architecture} & \textbf{Role in MLLMs} \\
\midrule
CLIP~\cite{radford2021learning}    & Image       & Dual-encoder contrastive (ViT + text Transformer) & Zero-shot vision encoder; image-text alignment \\
ViT~\cite{dosovitskiy2020image}    & Image       & Patch-based self-attention Transformer              & General-purpose image feature extractor \\
VAE~\cite{kingma2019introduction}  & Image       & Probabilistic encoder-decoder                       & Structured latent space for image generation \\
BLIP-2~\cite{li2023blip}          & Image       & Q-Former bridging module                            & Compact visual token extraction for frozen LLMs \\
Whisper~\cite{radford2023robust}   & Audio       & Seq-to-seq encoder-decoder (log-mel spectrogram)    & Multilingual speech recognition and transcription \\
WavLM~\cite{chen2022wavlm}        & Audio       & Self-supervised masked speech denoising Transformer  & Universal speech encoder for audio-text alignment \\
Flamingo~\cite{alayrac2022flamingo}& Video/Image & Perceiver Resampler + Gated Cross-Attention         & Few-shot visual reasoning across frames \\
TimeSformer~\cite{bertasius2021space}& Video     & Divided spatiotemporal attention Transformer         & Spatiotemporal feature extraction for video QA \\
Donut~\cite{kim2022ocr}           & Text-rich image & Swin Transformer encoder + BART-style decoder    & OCR-free document understanding \\
\bottomrule
\end{tabular}
\end{adjustbox}
\end{table}

\subsection{Fusion Strategy}
MLLMs integrate heterogeneous modalities through four main strategies, each making different trade-offs between cross-modal interaction depth and computational cost. \textit{Early fusion} merges raw or low-level features at the input stage before any independent unimodal processing, enabling the model to learn cross-modal dependencies from the first layer (e.g., Gemini~\cite{team2023gemini}), but requires full end-to-end training. \textit{Late fusion} preserves dedicated per-modality encoding pathways and combines outputs only at the prediction stage, offering modularity at the cost of fine-grained cross-modal correlation (e.g., Woodpecker~\cite{yin2024woodpecker}). \textit{Cross-modal attention fusion} establishes direct modality interactions throughout the architecture via attention mechanisms or gating functions, enabling richer contextual alignment (e.g., Multimodal-CoT~\cite{zhang2023multimodal}, BLIVA~\cite{hu2024bliva}). \textit{Hybrid fusion} combines two or more of the above strategies within a single model to balance localized and holistic reasoning (e.g., KOSMOS-2~\cite{peng2023kosmos}, GPT4RoI~\cite{zhang2023gpt4roi}). The detailed architectural trade-offs between these strategies are analyzed in Section~\ref{sec:mllm_tradeoffs}.

\subsection{Modality Approaches}
MLLMs integrate specialized encoder backbones to handle the distinct representation challenges of each modality. Image modalities rely on contrastive or patch-based vision models (e.g., CLIP~\cite{radford2021learning}, ViT~\cite{dosovitskiy2020image}) that project pixel data into embeddings compatible with LLM token spaces; generative image understanding additionally employs VAEs for structured latent representations~\cite{kingma2019introduction}. Audio modalities use sequence-to-sequence or self-supervised speech encoders that convert waveforms into temporal feature embeddings aligned with text. Video extends image encoding to the temporal dimension, requiring spatiotemporal models that aggregate frame-level features into compact token sequences. Text-rich images (documents, charts) employ either OCR pipelines for text extraction or OCR-free end-to-end document encoders that jointly model visual layout and embedded text. Table~\ref{tab:modality_approaches} summarizes representative encoder approaches across modalities.

\subsection{Comparative Analysis of MLLM Architectures}
\label{sec:mllm_comparison}

The rapid proliferation of Multimodal LLMs has produced architectures with substantially different design philosophies. Table~\ref{tab:mllm_comparison} compares six representative models across five dimensions: the vision–language fusion strategy, supported modalities, training approach, the key benchmark on which the model was evaluated, and representative benchmark performance, providing a basis for evaluating architectural trade-offs rather than merely cataloging approaches.

\textbf{Architectural trade-offs.} Three fusion strategies dominate the landscape: (1) \textit{Cross-attention injection} (Flamingo) inserts trainable cross-attention layers into a frozen LLM, preserving language model capacity at the cost of tight architectural coupling. (2) \textit{Bridging modules} such as the Q-Former (BLIP-2, InstructBLIP) project visual features into a compact set of learned query tokens before the LLM, enabling modular training stages but potentially losing fine-grained spatial information. (3) \textit{Linear projection connectors} (LLaVA) offer the simplest alignment and are highly efficient, but rely entirely on the richness of instruction tuning data for cross-modal alignment. Native end-to-end training (Gemini) achieves the strongest multimodal reasoning at the cost of substantially higher training resources and reduced modularity. For CAIS practitioners, the choice of fusion strategy determines not only benchmark performance but also the ease with which the vision component can be upgraded independently of the language backbone.

\subsection{Tasks}
MLLMs support a broad range of tasks spanning purely multimodal inference and compound applications that pair MLLMs with retrieval, reasoning, or control subsystems. Table~\ref{tab:mllm_tasks} summarizes the primary task categories with representative systems.

\begin{table*}[t]
\centering
\caption{Comparison of representative Multimodal LLM architectures across key design dimensions. Benchmark scores are taken from each model's original publication and reflect the specific model version evaluated at the time of release; later versions may differ.}
\label{tab:mllm_comparison}
\begin{adjustbox}{max width=\textwidth}
\begin{tabular}{lllllc}
\toprule
\textbf{Model} & \textbf{Fusion Strategy} & \textbf{Modalities} & \textbf{Training Approach} & \textbf{Key Benchmark} & \textbf{Score (\%)} \\
\midrule
Flamingo~\cite{alayrac2022flamingo}       & Cross-attention (frozen vision) & Image, video, text & Few-shot pretraining on image--text pairs & VQAv2 (fine-tuned) & 82.0 \\
BLIP-2~\cite{li2023blip}                 & Q-Former (bridging module)      & Image, text & Two-stage: Q-Former then LLM fine-tuning  & VQAv2 (fine-tuned) & 82.2 \\
LLaVA~\cite{liu2024visual}               & Linear projection connector     & Image, text & Instruction tuning on GPT-4 captions     & LLaVA-Bench (COCO) & 85.1 \\
InstructBLIP~\cite{dai2024instructblip}  & Q-Former (instruction-aware)    & Image, text & Instruction fine-tuning of BLIP-2         & ScienceQA & 90.7 \\
GPT-4V~\cite{achiam2023gpt}              & Proprietary (cross-modal attn.) & Image, text & RLHF + instruction tuning                & MMMU (val) & 56.8 \\
Gemini~\cite{team2023gemini}             & Native multimodal pretraining   & Image, audio, video, text & End-to-end multimodal training   & MMMU (Maj@32) & 62.4 \\
\bottomrule
\end{tabular}
\end{adjustbox}
\end{table*}

\subsection{Architectural Trade-offs in Multimodal Integration}
\label{sec:mllm_tradeoffs}

Three trade-off dimensions govern MLLM design. \textbf{Early vs.\ late fusion:} early fusion merges visual and textual tokens at input time, enabling deep cross-modal interaction at the cost of full end-to-end training (e.g., Gemini~\cite{team2023gemini}); late fusion processes modalities independently and merges only at the output, preserving modularity but losing fine-grained correlations. Most practical architectures use a frozen vision encoder that feeds compressed visual tokens into the LLM via a lightweight connector, balancing alignment quality with training efficiency. \textbf{Cross-attention injection vs.\ linear connectors:} cross-attention injection (Flamingo~\cite{alayrac2022flamingo}) enables dynamic visual attention but couples vision and language components architecturally; linear projection (LLaVA~\cite{liu2024visual}) is computationally negligible and decoupled but relies entirely on instruction-tuning data for alignment; Q-Former bridging (BLIP-2~\cite{li2023blip}) extracts a compact visual summary at the cost of a two-stage training procedure. \textbf{Instruction tuning vs.\ pretraining alignment:} full multimodal pretraining (Gemini, PaLM-E~\cite{driess2023palm}) achieves the strongest generalization; instruction tuning (LLaVA, InstructBLIP~\cite{dai2024instructblip}) is data-efficient but generalizes less robustly; RLHF (Fact-RLHF~\cite{sun2023aligning}) reduces hallucination but requires carefully designed multimodal reward models.

\subsection{Limitations and Failure Modes of MLLMs}
\label{sec:mllm_limitations}

Despite significant progress, MLLMs exhibit several systematic failure modes that are particularly important for CAIS designers to understand, as these failures propagate through downstream retrieval, agent reasoning, and orchestration components.

\textbf{Hallucination in visual grounding.} MLLMs frequently generate plausible-sounding but factually incorrect descriptions of visual content, a phenomenon distinct from text-only hallucination because it arises from misalignment between the visual embedding and the language model's prior. For example, models may confidently describe objects, attributes, or spatial relationships that are not present in the image. Woodpecker~\cite{yin2024woodpecker} proposes a post-hoc correction framework that identifies hallucinated visual claims and replaces them with grounded evidence, demonstrating that visual hallucination cannot be fully resolved by instruction tuning alone.

\textbf{Cross-modal inconsistency.} When the same query is posed in different modalities (e.g., as text vs.\ as a diagram), MLLMs may produce inconsistent or even contradictory answers, revealing that visual and textual representations are not truly unified in the model's internal state. This inconsistency is particularly problematic in CAIS contexts where a multimodal query triggers retrieval and downstream reasoning: an inconsistent grounding in the MLLM propagates errors through the entire pipeline.

\textbf{Evaluation fragility.} Standard multimodal benchmarks (e.g., VQAv2, SEED-Bench~\cite{li2023seed}, MM-Vet~\cite{yu2023mm}) primarily assess closed-ended visual question answering and do not capture open-ended generation quality, temporal reasoning over video, or robustness to adversarial visual perturbations. MM-SafetyBench~\cite{liu2024mm} highlights that models performing strongly on standard benchmarks can be easily jailbroken via multimodal adversarial inputs, underscoring the gap between benchmark performance and real-world robustness. For CAIS evaluation, this fragility implies that MLLM-component performance should be assessed with both standard benchmarks and adversarial/out-of-distribution test sets before integration into production pipelines.

Having examined the multimodal perception axis of CAIS, Section~\ref{sec:orchestration} turns to the fourth axis: the orchestration frameworks and coordination mechanisms that integrate RAG, agents, and MLLMs into coherent, production-ready pipelines.

\begin{table*}[t]
\centering
\renewcommand{\arraystretch}{1.2}
\caption{Representative MLLM task categories and example systems.}
\label{tab:mllm_tasks}
\begin{adjustbox}{max width=\textwidth}
\begin{tabular}{p{3.2cm}p{5.8cm}p{6.2cm}}
\toprule
\textbf{Task} & \textbf{Description} & \textbf{Example System} \\
\midrule
Multimodal Reasoning      & Infer relationships across modalities (e.g., visual QA, commonsense) & Socratic Models~\cite{zeng2022socratic}: zero-shot reasoning via prompted model composition \\
Cross-Modal Retrieval     & Map queries across modalities (text$\leftrightarrow$image) & GILL~\cite{koh2024generating}: frozen LLM + image encoder/decoder for retrieval \& generation \\
Content Generation        & Produce multimodal outputs (captions, image+text narratives) & Emu~\cite{sun2023generative}: interleaved text-and-image autoregressive generation \\
Contextual Understanding  & Ground language in visual context for detection/localization & ContextDET~\cite{zang2024contextual}: LLM outputs as prior for object detection \\
\midrule
+ CoT Prompting           & Decompose visual questions into sub-queries & DDCoT~\cite{zheng2023ddcot}: VQA model answers uncertainty-flagged sub-questions \\
+ RAG                     & Ground MLLM responses in retrieved multimodal evidence & RA-CM3~\cite{yasunaga2022retrieval}: CLIP-based dense retrieval into CM3 generator \\
+ RL Agents               & Use visual reward signals to align MLLM outputs & ESPER~\cite{yu2022multimodal}: CLIP cosine similarity as PPO reward \\
+ Embodied AI             & Reason over sensor data for navigation and manipulation & PaLM-E~\cite{driess2023palm}: robotic sensor tokens fused with language backbone \\
\bottomrule
\end{tabular}
\end{adjustbox}
\end{table*}
\section{Orchestration}
\label{sec:orchestration}

\begin{figure}[t]
    \centering
    \includegraphics[width=0.82\linewidth]{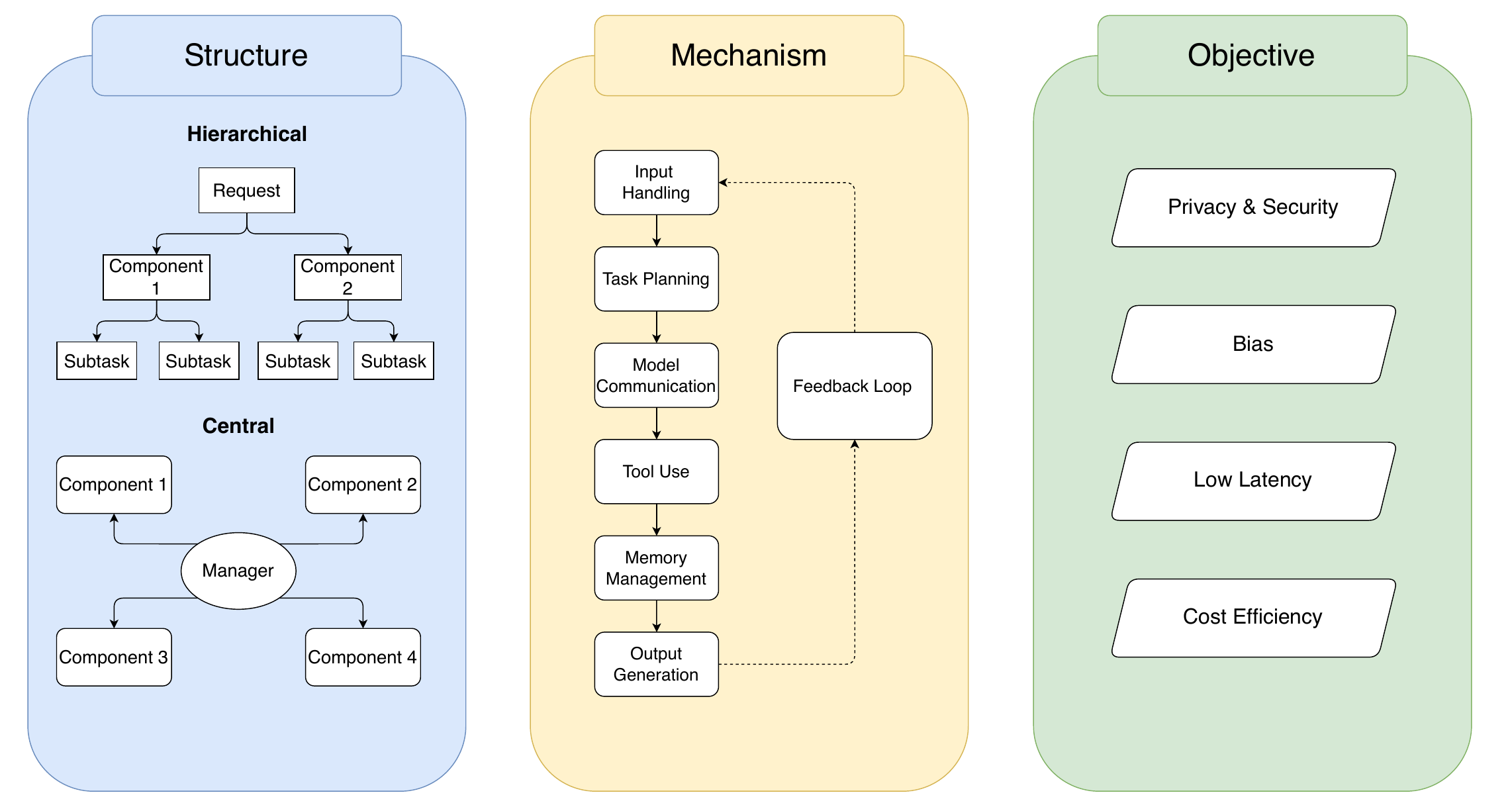}
    \caption{An overview of the Compound AI System orchestration layer, highlighting the relationship between its structure (e.g., hierarchical or centralized component organization), mechanism (such as input handling, task planning, and feedback loops), and objective (including privacy, low latency, and performance).}
    \Description{Diagram of the Compound AI System's orchestration layer showing how different structural types (hierarchical or centralized), mechanisms (input handling, task planning, feedback loops), and objectives (privacy, low latency, performance) interrelate.}
    \label{fig:orchestration}
\end{figure}

The orchestration of CAIS is inherently sophisticated, often requiring complex architectures. LLM-serving systems are designed to exceed the performance of standalone LLMs by integrating multiple layers, diverse models, or LLM-supported agents that collaborate to deliver optimal results. We categorize LLM-serving systems into three distinct layers: the structural layer, the mechanism layer, and the objective layer, as illustrated in Figure \ref{fig:orchestration}.

\subsection{Structural Layer}
The structural layer represents the architectural organization of components within a system, focusing on how tasks are distributed and coordinated. It includes two primary designs: hierarchical structure and centralized structure. This layer provides the foundational framework for the system's operational and interaction dynamics.

\subsubsection{Hierarchical Structure}
In a hierarchical structure, tasks are decomposed into subtasks managed by components organized in a tree-like hierarchy with clear dependencies. Components are modular, meaning they are distinct, self-contained, and interact with others in a coordinated manner. An example is MemGPT~\cite{packer2023memgpt}, which introduces a hierarchical memory system inspired by operating systems, comprising a main context (prompt tokens within the LLM) and external memory (for recall and archival storage), to address the limitations of LLMs' fixed context windows. The hierarchical structure has also been applied in data exploration~\cite{ma2023insightpilot}, multi-agent operating systems~\cite{mei2024llm}, and recommendation systems~\cite{wang2024llm4msr}.

\subsubsection{Centralized Structure}
In a centralized structure, a central manager oversees and coordinates interactions among components to ensure efficient collaboration. The central manager acts as a scheduler or coordinator, dispatching resources to the most suitable components based on the task.
For example, PagedAttention~\cite{kwon2023efficient} utilizes a centralized scheduler and distributed GPU workers, with a KV cache manager that dynamically allocates non-contiguous memory blocks for scalable LLM serving. The centralized structure has been applied to diverse domains including IoT~\cite{cui2024llmind} and multimodal systems~\cite{qin2024diffusiongpt}.

\subsection{Mechanism Layer}
The mechanism layer defines the operational processes that govern how a system handles tasks and produces results, ensuring seamless execution and adaptability. Besides input handling and output generation, it includes \textit{Task Planning}, which determines the steps required to achieve the desired outcome. For example, Infinite-LLM~\cite{lin2024infinite} integrates a distributed attention mechanism (DistAttention) with a centralized scheduler (gManager), decoupling attention computation from the main inference pipeline to enable flexible task allocation across GPU clusters.

\subsubsection{Model Communication}
\textit{Model Communication} enables different AI models or components within a system to exchange information. 
For example, TransLLaMa~\cite{koshkin2024transllama}, a simultaneous machine translation system, integrates an ASR model with a fine-tuned LLM. Tan et al.~\cite{tan2023large} propose a system involving multiple expert agents, each working with distinct abstraction spaces and combining primitive functions, structured prompting, and iterative feedback to refine outputs.

\subsubsection{Tool Use}
A key mechanism in CAIS is enabling LLMs to leverage external tools (e.g., APIs, search engines, databases, code executors)~\cite{qu2024tool, shen2024llm, wang2024tools}. This overcomes limitations such as static knowledge \cite{wen2024perception} and a lack of computational precision, allowing LLMs to interact with real-time data and perform specialized tasks. For instance, agents like OSAgent use standardized APIs to interact with operating systems~\cite{xu2024osagent}.

Tool use in LLM agents is generally enabled through two main approaches: training-based and prompt-based methods. \textit{Training-based methods} involve fine-tuning models to integrate tool use capabilities. Examples include GPT-4's structured function calling mechanism~\cite{achiam2023gpt} and Toolformer's self-supervised approach, which teaches the model to invoke tools such as search engines or translation services~\cite{schick2024toolformer}. In contrast, \textit{prompt-based methods} rely on strategic prompt design to guide LLM behavior. For instance, the ReAct framework interleaves reasoning with actions like tool invocation or web search via prompting, enabling complex task decomposition without additional fine-tuning~\cite{yao2023react}.

A specific application is enhancing LLMs with \textit{Web Search}. This addresses knowledge staleness and improves factuality. Models can be fine-tuned for web interaction, such as WebGPT using a text-based browser to search, navigate, and cite sources~\cite{nakano2022webgptbrowserassistedquestionansweringhuman}, or GopherCite using search to find supporting evidence for its claims~\cite{menick2022teachinglanguagemodelssupport}.

\subsubsection{Memory Management}
\textit{Memory Management} efficiently allocates and optimizes memory resources during tasks. For example, PagedAttention~\cite{kwon2023efficient} implements a paging-inspired memory management system that reduces waste and improves throughput by dynamically allocating non-contiguous blocks for key-value caches.

\subsubsection{Feedback Loop}
\textit{Feedback Loop} refers to a system mechanism in which outputs are cyclically returned as inputs to refine or optimize the process. For example, in Text-to-SQL systems~\cite{maamari2024end}, feedback loops refine SQL query generation based on execution results, improving success rates with each iteration.

\subsection{Comparative Analysis of Orchestration Frameworks}
\label{sec:orchestration_comparison}

A growing ecosystem of orchestration frameworks has emerged to simplify the construction and deployment of CAIS. These frameworks vary substantially in their architectural topology, coordination mechanism, tool support, and compatibility with emerging interoperability standards. Table~\ref{tab:orchestration_comparison} provides a structured comparison of five widely used frameworks.

\begin{table*}[t]
\centering
\renewcommand{\arraystretch}{1.2}
\caption{Comparison of CAIS orchestration frameworks across key design dimensions.}
\label{tab:orchestration_comparison}
\begin{adjustbox}{max width=\textwidth}
\begin{tabular}{llllcc}
\toprule
\textbf{Framework} & \textbf{Topology} & \textbf{Coordination Mechanism} & \textbf{Tool Support} & \textbf{Multi-Agent} & \textbf{MCP-Compatible} \\
\midrule
LangChain~\cite{langchain}        & Sequential / DAG   & Chain-based prompting, callbacks    & Extensive (150+ integrations) & Partial (via LangGraph) & Yes \\
AutoGen~\cite{wu2023autogen}      & Multi-agent graph  & Conversational agent messaging       & Moderate (code, search, APIs)  & Native              & Partial \\
DSPy~\cite{khattab2022demonstrate}& Pipeline (compiled)& Optimized prompt programs            & Limited (retrievers, LMs)      & No                  & No \\
LlamaIndex~\cite{llama_index}     & Hierarchical / DAG & Data connectors + query engine       & Extensive (data sources)       & Partial (via agents) & Yes \\
CrewAI~\cite{crewai2024}          & Role-based graph   & Task delegation between agents       & Moderate (tools via LangChain) & Native              & Partial \\
\bottomrule
\end{tabular}
\end{adjustbox}
\end{table*}

\textbf{Key trade-offs.} LangChain and LlamaIndex offer the broadest tool ecosystems and MCP compatibility, making them well-suited for general-purpose CAIS that require integration with diverse data sources and APIs. Their sequential/DAG topology is flexible but can become difficult to debug in complex pipelines. AutoGen and CrewAI are optimized for multi-agent coordination with native conversational or role-based messaging, but their tool ecosystems are narrower and MCP support is partial. DSPy takes a fundamentally different approach by compiling pipeline programs into optimized prompt sequences, trading runtime flexibility for reproducibility and systematic optimization—an advantage for research settings but limiting for production systems that require dynamic tool invocation. For CAIS practitioners, the choice of framework should be guided by: (1) whether multi-agent coordination is required, (2) the breadth of external tool integrations needed, and (3) whether standardized interoperability (MCP) is a deployment requirement.

\subsection{Objective Layer}
The objective layer governs the essential priorities of CAIS design: \textbf{Privacy and Security}—SECGPT~\cite{wu2024secgpt} uses isolated execution environments and hub-and-spoke models to protect against untrustworthy third-party apps, while Evertz et al.~\cite{evertz2024whispers} simulate attack scenarios to measure sensitive data leakage when LLMs interact with integrated tools; \textbf{Bias}—Sharma et al.~\cite{sharma2024generative} study how LLM-powered conversational search influences selective exposure and opinion polarization; \textbf{Low Latency}—LLM-Slice~\cite{liu2024llm} reduces transmission latency by dedicating wireless network slices for LLM task resource allocation; and \textbf{Cost-efficiency}—PALIMPZEST~\cite{liu2024declarative} compiles high-level task declarations into optimized execution plans that balance cost, runtime, and output quality.

\subsection{Standardization and Interoperability Protocols}
\label{sec:standardization}

A persistent challenge in Compound AI Systems is \textit{interoperability}: enabling heterogeneous LLMs, tools, retrievers, and agents to communicate reliably across framework and vendor boundaries. As CAIS pipelines grow in complexity, ad-hoc point-to-point integrations become increasingly brittle and difficult to maintain. This subsection surveys the emerging standardization landscape, organized as a three-layer stack.

\subsubsection{Schema Layer: Structured Tool Interfaces}
The lowest layer of interoperability standardizes how tools are \textit{described} to an LLM. OpenAI's function-calling specification (2023) introduced the now-prevalent convention of expressing tools as JSON-schema-described operations with typed arguments and return values. In 2024, OpenAI's Structured Outputs extension tightened this interface by enforcing strict JSON-schema adherence, reducing parsing failures in production systems. These schema conventions have become a de facto standard: most major LLM providers and orchestration frameworks now accept function-call specifications in a compatible format. However, schema standardization addresses only the description layer—it does not govern how tools are discovered, authenticated, or invoked across runtime boundaries.

\subsubsection{Tool-Access Protocol Layer: Model Context Protocol}
The \textit{Model Context Protocol} (MCP)~\cite{anthropic2024mcp,mcp2024spec}, introduced by Anthropic in November 2024, defines a universal open interface between AI systems (hosts and clients) and external tool servers. MCP adopts a host/client/server role model over JSON-RPC 2.0 and standardizes four core abstractions: \textit{resources} (data exposed by the server), \textit{prompts} (server-provided prompt templates), \textit{tools} (callable functions), and \textit{sampling} (server-initiated LLM requests). This design separates the AI application from the tool implementation, enabling a single MCP server (e.g., a database connector, a web search service) to be consumed by any MCP-compatible host without bespoke integration code.

\textit{Adoption and ecosystem growth.} Empirical studies of the MCP ecosystem show rapid adoption across both open-source and enterprise deployments~\cite{hasan2025mcp,guo2025mcp_ecosystem}. MCP is now natively supported by LangChain, LlamaIndex, and several cloud AI platforms, making it the most widely integrated tool-access protocol in the CAIS ecosystem. Mastouri et al.~\cite{mastouri2025openapi_mcp} demonstrate that existing REST APIs described in OpenAPI can be automatically compiled into MCP-compatible servers, allowing organizations to reuse their existing API infrastructure. Fei et al.~\cite{fei2025mcpzero} extend MCP with active tool discovery, enabling agents to retrieve relevant tools from large tool registries rather than consuming static schemas dumped into the context window.

\textit{Security limitations.} Despite its adoption, MCP introduces a standardized attack surface alongside its standardized interface. Radosevich and Halloran~\cite{radosevich2025mcp} demonstrate that MCP-enabled agents can be exploited to exfiltrate sensitive data, execute unauthorized actions, and bypass application-level safeguards. Hou et al.~\cite{hou2025mcp} provide a lifecycle-based threat model showing that vulnerabilities arise at server registration, capability negotiation, and tool execution stages, including privilege-separation and trust issues at ecosystem scale. These findings indicate that standardization of the interface layer does not automatically confer security: governance, authentication, and sandboxing must be co-designed with the protocol.

\subsubsection{Agent Interoperability Layer: Cross-Framework and Cross-Agent Protocols}
The highest layer of the standardization stack addresses interoperability not between an LLM and a single tool, but between \textit{whole agents, workflows, and ecosystems} running across organizational and runtime boundaries. Ehtesham et al.~\cite{ehtesham2025survey} survey four competing proposals operating at this layer:

\textbf{Agent-to-Agent (A2A)}, proposed by Google, enables structured communication between autonomous agents via a message-passing protocol that supports task delegation, capability advertisement, and result reporting.

\textbf{Agent Communication Protocol (ACP)}, developed by IBM, focuses on standardizing the message envelope and interaction semantics for multi-agent workflows within enterprise environments.

\textbf{Agent Network Protocol (ANP)} targets an ``agentic web'' vision in which agents discover one another via decentralized identity mechanisms and collaborate across open network boundaries~\cite{yang2025agent_protocols}.

\textbf{Agent Spec (Open Agent Specification)}~\cite{benajiba2025agentspec} proposes a framework-agnostic declarative representation for agents and workflows, with demonstrated portability across LangGraph, CrewAI, AutoGen, and other runtimes.

These proposals differ in scope: A2A and ACP target runtime message-passing; ANP targets open network discovery; Agent Spec targets static workflow portability. No single proposal currently addresses all interoperability tiers simultaneously.

\subsubsection{The Convergence-by-Layering Thesis}
A key insight from the 2024--2025 standardization literature is that interoperability in CAIS is converging not through a single universal standard, but through \textit{translation and layering}: OpenAPI-to-MCP compilers, protocol-agnostic tool registries, and shared agent representations allow older API standards, vendor schema conventions, and new agent protocols to coexist via adapters and intermediate representations~\cite{mastouri2025openapi_mcp,ehtesham2025survey}. For CAIS designers, this implies that near-term interoperability is achievable by adopting MCP for tool access, Agent Spec representations for workflow portability, and schema-level compatibility for LLM function calling, while the higher-level cross-agent protocol landscape continues to consolidate.

The following section reviews the benchmarks and evaluation metrics used to assess CAIS across all four axes surveyed in Sections~\ref{sec:rag}--\ref{sec:orchestration}.

\section{Benchmarks and Evaluation Metrics}
\label{sec:benchmarks_and_evaluation_metrics}

The rapid growth and advancement of CAIS have led to the development of numerous evaluation benchmarks and datasets. Because CAIS can have different objectives and structures, their respective evaluation benchmarks are diverse. This section introduces the key tasks, datasets, and quantitative metrics used for evaluating CAIS across the domains of RAG, Multimodal LLMs, LLM Agents, and Orchestration, as detailed in Table \ref{tab:evaluation_frameworks}.

% --- TABLE GOES HERE ---
\renewcommand{\arraystretch}{1.25}
\begin{table}[!htbp]
    \centering
    \caption{Summary of evaluation dimensions, benchmarks, and metrics for Compound AI Systems.}
    \resizebox{\textwidth}{!}
    {
    \begin{tabular}{p{2.5cm} p{2.5cm} p{5cm} p{4.5cm}}
        \toprule
        \textbf{Dimension} & \textbf{Task} & \textbf{Datasets/Benchmarks} & \textbf{Evaluation Metrics} \\
        \midrule
        % RAG PART
        \multirow{7}{*}{RAG} 
        & Open-domain QA & Natural Questions \cite{kwiatkowski2019natural}, TriviaQA \cite{joshi2017triviaqa}, WebQuestions \cite{berant2013semantic}, RealTimeQA \cite{kasai2023realtime} & Accuracy, F1 Score, EM \\
        & Passage Retrieval & MS MARCO \cite{bajaj2016ms}, BEIR \cite{thakur2021beir}, TREC-DL \cite{craswell2021trec} & MRR, nDCG, Precision, Recall, F1 Score \\
        & Knowledge QA & OpenBookQA \cite{mihaylov2018can}, PopQA \cite{mallen2022not}, TruthfulQA \cite{lin2021truthfulqa} & Accuracy, F1 Score, EM \\
        & Multi-hop QA & HotpotQA \cite{yang2018hotpotqa}, 2WikiMultihopQA \cite{ho2020constructing}, MuSiQue \cite{trivedi2022musique}, StrategyQA \cite{geva2021did} & EM, F1 Score, Accuracy \\
        & Extractive QA & SQuAD \cite{rajpurkar2016squad}, DROP \cite{dua2019drop} & EM, F1 Score, Precision, Recall \\
        & Summarization / Long-form Eval. & WikiAsp \cite{hayashi2021wikiasp}, Multi-News \cite{fabbri2019multi}, NarrativeQA \cite{kovcisky2018narrativeqa}, UniEval \cite{zhong2022towards} & ROUGE, BLEU, F1 Score \\
        & RAG Pipeline Evaluation & Ragas \cite{es2024ragas}, KILT \cite{petroni2020kilt} & Faithfulness, Context Precision, Context Recall, EM, F1 Score \\
        \midrule
        % Multimodal part
        \multirow{4}{*}{Multimodal LLM} 
        & Reasoning \& Commonsense QA & MM-Vet \cite{yu2023mm}, ALLaVA \cite{chen2024allava}, SEED-Bench \cite{li2023seed} & Accuracy, F1 Score, EM, Perplexity \\
        & Chart \& Document Understanding & ChartQA \cite{masry2022chartqa}, MMC-Benchmark \cite{liu2023mmc}, DocVQA \cite{mathew2021docvqa}, TextVQA \cite{singh2019towards} & Accuracy, F1 Score, BLEU, ROUGE \\
        & Safety \& Alignment & MM-SafetyBench \cite{liu2024mm} & Accuracy, F1 Score, Custom Safety Metrics \\
        & Multimodal Evaluation & MME \cite{fu2023mme}, MultiBench \cite{liang2021multibench} & Accuracy, F1 Score, MRR, Perplexity, BLEU, ROUGE \\
        \midrule
        % Orchestration part
        \multirow{3}{*}{\makecell[l]{Orchestration\\of CAIS}} 
        & Structural Layer & BigDataBench \cite{gao2018bigdatabench}, AIBench Training \cite{tang2021aibench} & Throughput, Latency Overhead, Bandwidth Utilization \\
        & Mechanism Layer & AISBench \cite{dong2025aisbench}, WebArena \cite{zhou2023webarena}, Long Range Arena \cite{tay2020long}, ZeroSCROLLS \cite{shaham2023zeroscrolls} & Mean Task Completion Time, Avg.\ API Response Time, Cache Hit Rate \\
        & Objective Layer & AI Fairness 360 \cite{bellamy2019ai} & Fairness-specific metrics \\
        \midrule
        % agent part
        \multirow{3}{*}{LLM Agents} 
        & Role-Playing & RoleLLM \cite{wang2023rolellm}, AgentBench \cite{liu2023agentbench}, AgentBoard \cite{ma2024agentboard} & Role-Consistency Score, Self-Correction Rate \\
        & Interactive Reasoning & AgentQuest \cite{gioacchini2024agentquest}, InfiAgent-DABench \cite{hu2024infiagent}, CriticBench \cite{lin2024criticbench} & Reasoning Trace Accuracy, Generalization Score \\
        & Tool Use & ML-Bench \cite{tang2023ml}, Berkeley Function Calling Leaderboard \cite{berkeley-function-calling-leaderboard} & Tool Call Accuracy, Token Efficiency \\
        \bottomrule
    \end{tabular}
    }
    \label{tab:evaluation_frameworks}
\end{table}
% --- END TABLE ---

\textbf{RAG evaluation} targets both retrieval quality (MRR, nDCG) and generation factuality (EM, F1, ROUGE, BLEU), with key benchmarks including HotpotQA~\cite{yang2018hotpotqa} for multi-hop reasoning, BEIR~\cite{thakur2021beir} for cross-domain retrieval generalization, and Ragas~\cite{es2024ragas} for end-to-end pipeline evaluation. \textbf{LLM agent evaluation} assesses role fidelity (Role-Consistency Score), reasoning quality (Reasoning Trace Accuracy), and tool-use efficiency (Tool Call Accuracy, Token Efficiency) in dynamic environments requiring planning, action, and self-correction. \textbf{MLLM evaluation} spans visual QA, chart and document understanding, and safety alignment, combining standard NLP metrics with specialized benchmarks such as MM-SafetyBench~\cite{liu2024mm} for adversarial robustness. \textbf{Orchestration evaluation} measures infrastructure performance (throughput, latency), system coordination (task completion time, cache hit rate), and alignment with fairness objectives across the structural, mechanism, and objective layers. Full benchmark and dataset listings for all four dimensions are provided in Table~\ref{tab:evaluation_frameworks}. Persistent gaps—particularly the lack of unified system-level metrics and robust multimodal benchmarks—motivate the challenges and open directions discussed in the following section.

\section{Challenges, Limitations, and Opportunities}  
\label{sec:challenges_limitations_and_opportunities}

\subsection{Challenges and Limitations}
Compound AI Systems represent a significant advancement in the capabilities of LLM-based architectures, yet they are not without challenges and limitations. As this field matures, it is crucial to understand the key obstacles that hinder its scalability, efficiency, and safety, while also identifying the most promising future directions.

\subsubsection{System Complexity and Scalability} Compound AI Systems often involve the integration of multiple components—retrievers, agents, multimodal encoders, memory modules, and orchestration mechanisms. This architectural complexity introduces engineering overhead, increasing the difficulty of deployment, debugging, and system optimization. Moreover, the added complexity can lead to performance bottlenecks and increased inference latency, particularly when coordinating multiple LLMs and tools in real-time.

\subsubsection{Evaluation and Benchmarking} Evaluating Compound AI Systems is inherently difficult due to the diversity of their components and the dynamic nature of their outputs. Traditional NLP benchmarks do not adequately capture the interactive or multimodal capabilities of such systems. There is a lack of unified evaluation frameworks that consider system-level performance metrics, such as latency, robustness, and resource usage, alongside task-specific accuracy.

\subsubsection{Tool and Component Integration} Despite advances in tool-use and agent-based planning, seamless integration between LLMs and external components remains a challenge. Issues such as API misalignment, inconsistent data formats, limited error handling, and brittle tool chaining cause failures in real-world applications. Moreover, many tools and retrievers were not originally designed for interaction with LLMs, leading to interoperability and reliability issues.

\subsubsection{Standardization and Ecosystem Fragmentation} The rapid growth of the CAIS ecosystem has produced a fragmented landscape of incompatible frameworks, tool schemas, and agent protocols. While emerging standards such as the Model Context Protocol (MCP)~\cite{anthropic2024mcp} and Agent Spec~\cite{benajiba2025agentspec} address parts of this problem (as discussed in Section~\ref{sec:standardization}), no single standard currently covers the full interoperability stack from tool schemas through agent communication to cross-framework workflow portability. Security governance of standardized interfaces is also under-developed: studies have demonstrated that MCP's open extensibility introduces privilege-separation vulnerabilities at ecosystem scale~\cite{radosevich2025mcp,hou2025mcp}. Developing governance models, authentication mechanisms, and sandboxing standards that co-evolve with protocol design is a critical open challenge for the field.

\subsubsection{Multimodal Alignment and Evaluation} Despite rapid progress in MLLMs, cross-modal consistency and evaluation robustness remain unsolved. As shown in Section~\ref{sec:mllm_limitations}, MLLMs exhibit visual hallucination, cross-modal inconsistency, and fragility to adversarial visual inputs that are not captured by standard benchmarks. Developing evaluation frameworks that assess multimodal CAIS components under realistic, out-of-distribution, and adversarial conditions is an important open problem.

\subsection{Opportunities}

The challenges above delineate the most fertile frontiers for the next generation of CAIS.

\subsubsection{Unified and Modular Architectures}
Designing CAIS with clean interfaces between retrievers, agents, encoders, and orchestration layers—so components can be swapped without cascading side effects—is a key priority. Standardized protocols (MCP, A2A; Section~\ref{sec:standardization}) are early steps, but interoperability benchmarks and certified adapter libraries are still needed.

\subsubsection{End-to-End Trainable Compound Pipelines}
Current CAIS are assembled from independently pretrained modules with no joint training objective; end-to-end differentiable or RL-based training of retriever--generator--agent pipelines~\cite{lin2024llm,lee2025compound} is a promising direction. Key open challenges are credit assignment across heterogeneous components and fine-tuning strategies that avoid catastrophic forgetting.

\subsubsection{Multimodal Grounding and Evaluation}
Advances in cross-modal alignment for video, audio, and 3D inputs will expand the task scope of CAIS. Future benchmarks should probe compositional reasoning, temporal consistency, and adversarial robustness—going beyond aggregate accuracy.

\subsubsection{Self-Adaptive and Meta-Agent Orchestration}
Static workflow graphs are brittle in open-ended environments; self-adaptive orchestration that monitors performance, reconfigures components, and allocates resources dynamically is an important frontier. Meta-agent architectures—where a high-level planner supervises specialist sub-agents—offer a path toward more robust and explainable CAIS.

\subsubsection{Human-AI Collaboration and Interpretability}
In high-stakes domains (medicine, law, finance), explaining compound reasoning chains—tracing outputs back through orchestration, retrieval, and tool calls—is essential. Interactive explanation interfaces, uncertainty quantification, and controllable planning mechanisms for domain-expert oversight are open research opportunities.

\subsubsection{Sustainable and Efficient Deployment}
The multi-component nature of CAIS amplifies inference cost; speculative retrieval, cached reasoning, and adaptive computation (early-exit agents) are promising directions for reducing this overhead. Efficiency benchmarks measuring system-level throughput and cost per task—rather than per-component metrics—would accelerate progress.

\section{Conclusion}
\label{sec:conclusion}

This survey has presented a systematic, systems-level synthesis of Compound AI Systems (CAIS)—the emerging paradigm in which large language models are augmented with external components to overcome the inherent limitations of standalone models. Drawing on a structured search of over 220 papers published between 2020 and early 2026, we have organized the CAIS landscape along four foundational axes: Retrieval-Augmented Generation (RAG), LLM Agents, Multimodal LLMs (MLLMs), and Orchestration.

\textbf{Key contributions.} For each axis, we have provided both descriptive coverage of representative systems and comparative analysis of design trade-offs through dedicated comparison tables (Tables~\ref{tab:rag_comparison}--\ref{tab:orchestration_comparison}). We introduced a unified cross-axis pipeline model (Figure~\ref{fig:unified_pipeline}) and formal instantiation in terms of $f(L, C, D)$ that shows how the four axes co-constrain one another in a complete CAIS. We surveyed the emerging standardization landscape—MCP, A2A, ACP, ANP, and Agent Spec—and argued that interoperability is currently converging through translation and layering rather than a single universal standard. We also deepened the comparative treatment of MLLM architectures and identified three systematic failure modes—visual hallucination, cross-modal inconsistency, and evaluation fragility—that are particularly consequential for CAIS integration.

\textbf{Open challenges.} Despite this progress, significant challenges remain. System complexity and debugging difficulty grow super-linearly with the number of integrated components. Evaluation frameworks lag behind system capabilities, particularly for interactive, multimodal, and long-horizon tasks. Standardization and security governance of tool-access and agent communication protocols are still immature. Cross-modal alignment in MLLMs also remains an unsolved problem at the intersection of representation learning and evaluation methodology.

\textbf{Looking forward.} We anticipate that the most impactful near-term advances in CAIS will come from three directions: (1) unified, modular architectures that allow components to be upgraded or replaced without retraining adjacent modules; (2) end-to-end trainable pipelines in which retrievers, agents, and generators are jointly optimized; and (3) principled governance frameworks for standardized interoperability that address both portability and security. As CAIS continue to expand into real-world, multimodal, and interactive environments, the systems-level perspective developed in this survey—synthesizing retrieval, agency, perception, and orchestration into a coherent analytical framework—will be essential for designing, evaluating, and advancing the next generation of intelligent applications.

% \begin{acks}
% \end{acks}

\bibliographystyle{ACM-Reference-Format}
\bibliography{sources}

\appendix

\end{document}